\documentclass[useAMS,usenatbib]{mn2e}
\usepackage{latexsym,hhline,epsfig,longtable}
\usepackage{graphicx,color}

\def\iso#1{$^{#1}$} 
\def\msun{M$_\odot$}
\usepackage{color} 
\usepackage[normalem]{ulem}                    
\def\iso#1{$^{#1}$}
{}

\title[Partial mixing and the formation of \iso{13}C pockets]
{Partial mixing and the formation of \iso{13}C pockets in AGB stars: 
effects on the $s$-process elements}
\author[J. F. Buntain et al]{J. F. Buntain$^{1}$, C. L. Doherty$^{2,1}$, 
M. Lugaro$^{2,1}$, J. C. Lattanzio$^{1}$, 
R. J. Stancliffe$^{3}$ \newauthor and A. I. Karakas$^{1,4}$\\ 
\\
$^{1}$Monash Centre for Astrophysics (MoCA), School of Physics and Astronomy, Monash 
University, Victoria 3800, Australia\\
$^{2}$Konkoly Observatory, Research Centre for
Astronomy and Earth Sciences, Hungarian Academy of Sciences, H-1121
Budapest, Hungary\\
$^{3}$Argelander Institute for Astronomy, University of Bonn, Auf dem Huegel 71, D-53121 
Bonn, Germany \\
$^{4}$Research School of Astronomy and Astrophysics, Australian National University, 
Canberra, ACT 2611, Australia.}

\begin{document} 
\maketitle

\begin{abstract}
The production of the elements heavier than iron via $slow$ neutron captures (the $s$ process)
is a main feature of the contribution of asymptotic giant branch (AGB) stars of low mass ($<$ 5 Msun) to
the chemistry of the cosmos. However, our understanding of the main neutron source, the
\iso{13}C($\alpha$,n)\iso{16}O reaction, is still incomplete. It is commonly assumed that in AGB stars
mixing beyond convective borders drives the formation of \iso{13}C $pockets$. However, there is
no agreement on the nature of such mixing and free parameters are present. By means of a parametric model
we investigate the impact of different mixing functions on the final $s$-process abundances in
low-mass AGB models. Typically, changing the shape of the mixing function or the mass extent of the
region affected by the mixing produce the same results. Variations in the relative abundance distribution
of the three $s$-process peaks (Sr, Ba, and Pb) are generally within +/-0.2 dex, similar to the
observational error bars. We conclude that other stellar uncertainties - the effect of rotation and of
overshoot into the C-O core - play a more important role than the details of the mixing function. The
exception is at low metallicity, where the Pb abundance is significantly affected. In relation to the
composition observed in stardust SiC grains from AGB stars, the models are relatively close to the data
only when assuming the most extreme variation in the mixing profile.
\end{abstract}

\section{Introduction}\label{sec:Introduction}

Asymptotic giant branch (AGB) stars are a significant site of the origin of chemical matter 
in the Universe, and in particular of the elements heavier than Fe. 
The AGB is populated by low- and intermediate-mass stars (in the range 
$\sim$ 0.5 to 10 \msun) evolved past core H and He burning.
Their structure is characterised by an inert, degenerate C-O core,  
a He- and a H-burning shell, separated by a He-rich ``intershell'', and 
an extended, convective, H-rich envelope. The H-burning 
shell is active most of the time, except when recurrent thermal instabilities of the He 
shell (thermal pulses, TPs) result in partial He burning and the $^{12}$C enrichment of the 
intershell. 
Most of the stellar mass is located in the envelope, which is  
is eroded by strong, dusty stellar winds. 
Recurrent dredge-up episodes (the third dredge-up, TDU) that may follow each TP bring 
material rich in nuclei freshly synthesized by nuclear reactions,
from the intershell to the surface of the star. 
From there, the winds carry the enriched 
matter into the stellar surroundings. 
Once the winds have removed most 
of the envelope mass, the star moves onto the post-AGB track, evolving 
at constant luminosity towards 
higher temperatures. The C-O core eventually is left as a cooling white dwarf. 
For detailed reviews on AGB stars see 
\citet{herwig05}, \citet{straniero06}, and \citet{karakas14}.

The cosmic abundances of the elements heavier than Fe are predominately produced via the 
capture of free neutrons because their large number of protons ($>$26) results in a strong 
Coulomb barrier \citep{b2fh}. In the case of the $slow$ neutron-capture process (the 
$s$ process) typical neutron 
densities are of the order of 10$^7$ cm$^{-3}$, at which 
the timescale of the neutron capture is  
longer than the timescale of the $\beta$ decay of unstable isotopes and the nuclear
production path proceeds along a chain of stable nuclei. Nuclei with magic numbers of 
neutrons $N$ along the path, e.g., \iso{88}Sr ($N=50$), \iso{138}Ba ($N=82$), 
and \iso{208}Pb ($N=126$), have lower neutron-capture 
cross sections relative to nearby nuclei and tend to accumulate during the $s$ process.
This leads to the presence of three peaks in the $s$-process abundance distribution, 
corresponding to the three magic numbers above. 
The $s$-process stellar sites are identified with  
the hydrostatic burning phases of massive 
stars \citep{raiteri91,pignatari10} and the AGB phase \citep{iben78,gallino98}.

In AGB stars the $s$ process occurs in the intershell, both within the convective regions 
generated by the TPs and during the periods between TPs (the so-called interpulse) in radiative 
conditions 
\citep{straniero97,gallino98,busso99,goriely00,lugaro03a,cristallo09,bisterzo10,lugaro12}. In 
the convective regions the neutron source is the $^{22}$Ne($\alpha$,n)$^{25}$Mg reaction. 
Nuclei of $^{22}$Ne are abundant inside the convective TPs because they are the result of double 
$\alpha$ captures on the $^{14}$N present in the H-burning ashes and ingested by the TP 
convective zone. 
Neutrons are released over a relatively short time scale (of the order of years), which results in 
relatively high neutron densities \citep[approximately, up to 10$^{10}$\,cm$^{-3}$ in AGB 
stars of initial mass $<$ 4 \msun\ and to 10$^{13}$\,cm$^{-3}$ in AGB stars of higher 
masses, e.g.][]{vanraai12,fishlock14,straniero14}.
Such neutron densities drive the activation of 
{\em branching points} on the $s$-process path, where unstable nuclei with half lives of the 
order of or larger than a day can capture a neutron instead of decaying, resulting in a branch 
on the path of neutron captures. However, the $^{22}$Ne neutron source requires 
temperatures above 300 MK to be efficiently activated, which are only found in stellar models 
of mass above around 3 \msun\ \citep{iben78,abia01,vanraai12,karakas12}. 

In the lower-mass AGB stars, which are observed to be strongly $s$-process enriched 
\citep[e.g.,][]{busso01,abia02}, the best candidate neutron source is instead the 
\iso{13}C($\alpha$,n)\iso{16}O reaction, which is activated at 90 MK. 
Because the amount of $^{13}$C produced by H burning is not 
enough to explain the observed \emph{s}-process enhancements, 
it is assumed that a partial mixing zone (PMZ) forms at the deepest extent of each TDU episode, 
where protons from the envelope are mixed into the intershell.
The resulting compositional profiles are mostly determined by the relative rates of the 
proton-capture reactions on \iso{12}C and \iso{13}C, which are functions of the number of protons, 
\iso{12}C, and \iso{13}C.
In the region of the PMZ where the ratio of the number of protons to $^{12}$C is $<$0.5 the 
$^{12}$C(p,$\gamma$)$^{13}$N($\beta^{+}$ $\nu$)$^{13}$C reaction chain results
in a thin region rich in $^{13}$C known as the $^{13}$C {\it pocket} \citep{goriely00,lugaro03a}.
Typically, an extent in mass of the pocket 
$\sim 10^{-3,-4}$ \msun\ is required to match the observations \citep{gallino98}.
Where the number of protons is $>$0.5 further $^{13}$C+p reactions produce 
a $^{14}$N-rich region, which we will refer to as the $^{14}$N pocket.

The $^{13}$C nuclei typically burn over long timescales ($\sim 10^4$ yr) 
during the interpulse periods, which results in a slow burning and  
low neutron densities, roughly
10$^{7}$\,cm$^{-3}$. Because the $^{13}$C nuclei are produced  
from proton captures on mostly primary $^{12}$C made by the triple-$\alpha$ reaction,
the $^{13}$C neutron source is of primary origin, i.e., 
the number of $^{13}$C nuclei is largely independent of 
the stellar metallicity. The number of Fe seeds is instead 
metallicity dependent. It follows that the number of free neutrons available in the $^{13}$C 
pocket increases with decreasing metallicity and that the second and third $s$-process 
peaks at Ba and Pb are more efficiently produced as the the metallicity decreases. 
This effect explains the origin and evolution of most of 
of the $s$-process elements in our Galaxy \citep{travaglio99,travaglio01a,travaglio04}. 

The main problem with the $^{13}$C-pocket model is that 
although many possibilities have been proposed, 
the main mixing mechanism leading to the formation of the $^{13}$C pocket
is still unknown \citep[see, e.g., discussion in][]{busso99,karakas14} 
The aim of the present paper is to explore a variety of mixing profiles
and their impact on $s$-process abundances considering the full evolution along the thermally
pulsing AGB phase of six stellar models of masses between 1.25 and 4 \msun\footnote{We do not consider 
higher masses because observations 
of AGB stars indicate 
that the formation 
of the $^{13}$C pocket does not occur, or that its impact is very marginal, 
in more massive AGB stars where the $^{22}$Ne neutron source is activated instead
\citep{garcia13}.
This is in agreement with the results obtained by   
models that include the PMZ using overshoot 
\citep{goriely04,cristallo15a}.}
\msun\ and metallicities 0.02, 0.01, and 0.0001.
This exploration will reveal to us if and when significant variations 
in the final $s$-process abundances occur due to changes in mixing profile leading to the 
formation of the PMZ.
In Sec.~2 we summarise the different scenarios and models 
for the formation of the \iso{13}C pocket currently present in the literature. 
In Sec.~3 we present our models, in Sec.~4 the results, and 
Sec.~5 is dedicated to discussion and conclusion.

\section{Current models for the formation of \iso{13}C pockets} 

A range of possibilities and models has been proposed so far to mix protons into the He intershell
at the deepest extent of each TDU episode, where a sharp discontinuity is present 
between the H-rich convective envelope and the He-rich radiative interhell.
Early on, \citet{hollowell88} showed that in AGB stars of low mass and low metallicity
a semi-convective region can form between the H-rich
envelope and the C-rich intershell after each TP due to the increased opacity below 
the convective envelope. More recently, \cite{herwig00} and 
\cite{cristallo09} successfully employed 
convective overshoot to reproduce the mixing mechanism 
responsible for the PMZ. \cite{herwig00} modelled the overshoot via the 
diffusion coefficient while \cite{cristallo09} via the 
convective velocity, both using an exponential decay function of the form: 

 \begin{equation}
 A \simeq A_{0}\, \exp \left(-\frac{d}{\beta H_{p}}\right)
 \label{eq:4.2}
 \end{equation}

where A is either the diffusion coefficient or the convective velocity, $A_{0}$ its value close to the 
formal convective boundary (defined by the
Schwarzschild criterion), $d$ the distance from the formal convective boundary,  
H$_{p}$ the pressure scale height, and $\beta$ (also named $f$) the free overshoot parameter
that controls the profile of the exponential decay, i.e., the region affected by the mixing and 
the extent of the final $^{13}$C pocket.

\citet{cristallo09,cristallo11,cristallo15a} presented and discussed 
detailed $s$-process models and 
results based on this scheme (the FRUITY database, http://fruity.oa-teramo.inaf.it/). 
The results are very similar to those obtained using the parametric model that we present below, 
where the PMZ is introduced artificially using an exponential mixing profile 
\citep[for a detailed comparison between the two sets of models 
see][]{lugaro12,fishlock14,karakas16}. 

The main difference between \cite{herwig00} and \cite{cristallo09} is that  
\cite{cristallo09} only apply the overshoot to the base
of the convective envelope during the TDU, while \cite{herwig00} applies it to all convective 
boundaries, including those of the TPs. This 
results in the dredge-up of C and O from the core 
into the intershell with 
crucial implications on the formation of the $^{13}$C pocket and the 
final $s$-process results \citep{lugaro03a}. In fact, a higher amount 
of $^{12}$C in the intershell leads to a higher abundance of $^{13}$C, and subsequently 
a higher number of free neutrons. Full $s$-process models including this effect
have been developed by \citet{pignatari16} and \citet{battino16}.

Internal gravity waves caused by convective motions beating on
the convective/radiative interface are also suggested 
by \citet{denissenkov03a,battino16} to produce the PMZ and produce PMZ similar to those resulting 
from the overshoot models.
 
Another scenario is related to 
mixing driven by magnetic fields, as discussed in detail by \citet{nucci14} and 
\citet{trippella16}. The results from this scenario differs from all the others in that 
the mixing mechanism is not tied to the convective envelope boundary, with the result that  
the $^{13}$C pocket is more extended in mass and the $^{14}$N pocket does not form.

Finally, mixing resulting from  
stellar rotation produce $^{13}$C pockets too small ($\sim 10^{-7}$ \msun) to 
reproduce the observations \citep{langer99}. Furthermore,
rotational mixing can be present while the neutrons are released inside 
$^{13}$C and $^{14}$N pockets produced, e.g., by overshoot.
This mixing can strongly inhibit the $s$ process by carrying
$^{14}$N, which is a neutron poison via the $^{14}$N(n,p)$^{14}$C reaction \citep{wallner16}, 
from the 
$^{14}$N pocket (and/or the H-burning ashes, if the $^{14}$N pocket is absent) 
into the $^{13}$C pocket \citep{herwig03,siess04,piersanti13}.

In a number of the current models of the $s$ process in AGB stars, 
a PMZ or a $^{13}$C pocket is introduced artificially at the deepest 
extent of each TDU episode, with its  
extension and mixing or abundance profile treated as 
relatively free parameters. The models of \citet{gallino98} 
\citep[and follow-up studies by, e.g.,][]{bisterzo10,bisterzo11,liu15} are calculated by
including some $^{13}$C in the intershell and exploring a large variety
of extents and abundance profiles. The models of, e.g.,  
\citet{goriely00,lugaro12,fishlock14,shingles15,karakas16} instead are calculated 
including a PMZ by means of an artificial mixing profile, driving the 
mixing of protons that subsequently produce a $^{13}$C abundance profile. 
The mixing is modelled using an exponential function where the exponent 
is a linear function of the mass (``standard'' case, see details in Sec.~2.1)
and different mass 
extents 
of the mixing are tested, from about 1/100th to 1/5th of the mass of the intershell.

\cite{goriely00} further analysed the effect of modifying the mixing function to
generate proton profiles decreasing with the mass depth faster or slower than the standard 
case.
They found that changing the H profile does not significantly 
affect the final $s$-process abundance distribution and concluded that these
are only marginally dependent on the shape of the H profile, 
whereas the extent in mass of the mixing and the amount of 
TDU more significantly affect the surface 
enrichment. 
However, the study carried out by \citet{goriely00} was quite limited. The
nucleosynthesis was followed 
in four stellar models of masses between 1.5 and 3 \msun\ and metallicities between 0.001 and 0.02   
only during one representative interpulse and TP phase, 
rather than over the whole AGB phase. Potential feedback effects on the nucleosynthesis 
during the AGB evolution 
were not considered. Furthermore, given the large variety of possible mixing
profiles potentially 
resulting from the different physical processes for the formation of the PMZ described above,
a detailed study dedicated to 
the impact of such variations 
on the final $s$-process abundances is required and presented here.

\section{Stellar Models}
\subsection{The stellar structure sequences}

We use stellar structure models previously calculated from the zero-age main 
sequence to the end of the AGB phase using the Monash stellar 
structure code \citep{lattanzio86} and including mass loss during the AGB 
phase using the prescription of \citet{vw93}. We considered the
1.5\,M$_\odot$ \emph{Z} = 0.0001 model from \citet{lugaro12}, the 3 
\msun\ and 4 \msun\ models of metallicity $Z=0.02$ from 
\cite{karakas10a}, the 3 \msun\ model of $Z=0.01$ from 
\citet{shingles13}, and the 1.25 \msun\ and 1.8 \msun\ models of 
metallicity $Z=0.01$ from \citet{karakas10b}. The inputs used for the structure 
calculations were the same for all the models, except that  
the $Z=0.01$ models were 
computed with the inclusion of the C- and N-rich low temperature opacity 
tables from \citet{lederer09}, and that convective overshoot was 
included in the 1.25 \msun\ and 
1.8 \msun\ models by extending the position of 
the base of the convective envelope by $N_{\rm ov}$ pressure-scale 
heights with $N_{\rm ov} = 4$ and 3, respectively 
\citep[for more details, see][]{karakas10b}. 
This overshoot has the effect of deepening the TDU, but 
 does not lead to the formation of the PMZ in the top layers of the 
 intershell because we use homogeneous mixing in the overshoot region.

The main structural features of the selected models are presented in 
Table~\ref{table:models} where we report: the initial stellar mass 
(Mass) and metallicity ($Z$), the number of thermal pulses (TPs), 
the number of TDU episodes (TDUs), 
the total mass dredged-up by the TDU ($M^{\rm tot}_{\rm dred}$), the 
maximum temperature in the TPs ($T^{\rm{max}}_{\rm{TP}}$), and the final envelope mass 
($M^{\rm{fin}}_{\rm{env}}$). The values of $M^{\rm{fin}}_{\rm{env}}$ 
in the case of the $Z=0.02$ models are still relatively high due 
to convergence issues at the end of the calculations \citep{lau12}. Due to the high mass-loss rate at 
this point of the evolution, only one or two more TPs are possible before the envelope is
lost and it is uncertain if any more TDUs would occur following these TPs. 
Hence, we consider the final abundances 
computed for these models close enough to those that would result from adding 
an extra one or two TPs.

 \begin{table}
 \centering
 \caption{Details of the stellar structure models. All the masses are in unit of \msun\ and 
the temperature in MK.} 
 \label{table:models}
 \begin{tabular}{lllllll}
 \hline 
 Mass & \emph{Z} & TPs & TDUs &
 $M^{\rm tot}_{\rm dred}$ & $T^{\rm max}_{\rm TP}$ & 
 $M^{\rm fin}_{\rm env}$ \\  
 \hline 
 1.25 & 0.01 & 10 & 3 & 0.013 & 246 & 0.026  \\  
 1.8 & 0.01 & 12 & 6 & 0.041 & 266 & 0.014   \\ 
 3 & 0.01 & 22 & 17 & 0.120 & 306 & 0.004   \\ 
 3 & 0.02 & 25 & 16 & 0.081 & 302 & 0.676  \\ 
 4 & 0.02 & 17 & 15 & 0.056 & 332 & 0.958  \\ 
 1.5 & 0.0001 & 18 & 15 & 0.059 & 282 & 0.022   \\
 \hline
 \end{tabular}
 \end{table}

\subsection{The stellar nucleosynthesis sequences}
\label{sec:nucleo}

 The $s$-process 
 nucleosynthesis was calculated using the Monash post-processing code \citep{cannon93}, 
 which takes stellar structure information, such as temperature, density, and 
 convective velocity as a function of interior mass and time, and
 solves implicitly the set of equations that simultaneously describe the
 abundance changes due to nuclear reactions and 
to mixing via a two-stream advective scheme.
We assumed scaled-solar initial compositions, taking the solar
 abundances from the compilation by \citet{asplund09}.
 We include the PMZ artificially in the 
 post-processing phase by forcing the code to mix a small amount of 
 protons from the envelope into the intershell at the end of each TDU.
 The method is described in detail in the following section.
Here we note that while this method allows us to investigate 
feedback from the inclusion of the PMZ on the nucleosynthesis, 
we cannot investigate the potential feedback on the 
stellar structure, given that the PMZ is included only in the post-processing. 
This needs to be kept in mind particularly in relation to the 
case when the \iso{13}C pocket is ingested in the following TP (as detailed in the 
next section), which 
could lead to modification of the TP structure \citep{bazan93}. 

 We employed a network of 320 nuclear species from neutrons and protons up 
 to bismuth. Nuclear reaction rates were included 
 using the $reaclib$ file provided by the Joint 
 Institute for Nuclear Astrophysics \cite[JINA,][]{cyburt10}, as of May 
 2012 (reaclib\_V2.0). The rates of the neutron source reactions correspond to 
 \citet{heil08} for the \iso{13}C($\alpha$,n)\iso{16}O and to 
 \citet{iliadis10} for the \iso{22}Ne($\alpha$,n)\iso{25}Mg and 
 \iso{22}Ne($\alpha$,$\gamma$)\iso{26}Mg reactions. For the  
 neutron-capture cross sections, the JINA reaclib database includes 
 the KADoNiS database \citep{dillmann06}\footnote{We used 
 the rates labelled as $ka02$ in the JINA
 database (instead of $kd02$) as they provide the best fits to KADoNiS 
 at the temperature of interest for AGB 
 stars.}. 

\subsubsection{The inclusion of the PMZ}

We test the effect of different types of PMZs on the resulting $^{13}$C 
pockets and the ensuing nucleosynthesis. 
Our abundance profiles within the PMZ are constructed in the following way. 
As commonly assumed, we postulate that the PMZ 
is the result of mixing occurring below the bottom of the convective 
envelope during each TDU episode 
at the point in time when the envelope reaches its 
maximum inward penetration in mass. We define that the mixing
extends for a depth $M_{\rm PMZ}$, in other words, that it 
reaches down to a point in mass of $M_{\rm PMZ}$ below the base of the
convective envelope. To describe the mixing function over mass
we scale the mass coordinate within the PMZ to a variable \emph{m$_{\rm scaled}$} 
that varies between 0 and 1 over the PMZ:

 \begin{equation}
 m_{\rm scaled} = \frac{m - m_{\rm base}}{m_{\rm top} - m_{\rm base}}
 \label{m_{sc}}
 \end{equation}

where \emph{m} is the mass variable within the PMZ, $m_{\rm top}$ is the mass at the 
top of the PMZ (i.e., the mass coordinate of the base of the convective envelope) and 
$m_{\rm base}$ is the mass at the base of the PMZ (i.e., \emph{m$_{\rm base}$} = \emph{m$_{\rm top}$} $-$ 
$M_{\rm PMZ}$). 

The mixing function is defined as the exponential $10^{\emph{f}}$, 
where in the ``standard'' case \emph{f} is a linear function of the mass:

 \begin{equation}
 f = 4\,m_{\rm scaled} - 4
 \end{equation}

\begin{figure}
 \centering
 \includegraphics[width=8.5cm,angle=0]{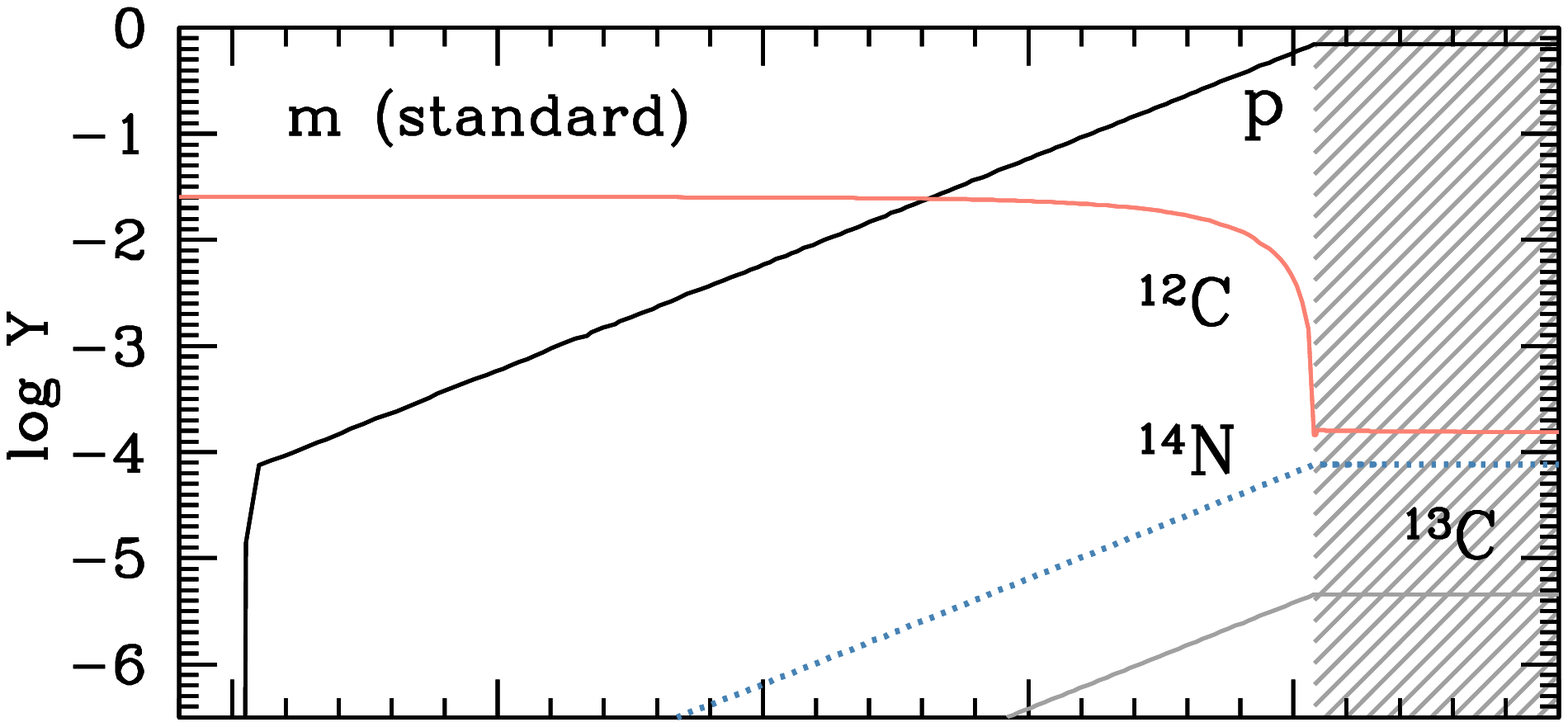} 
 \includegraphics[width=8.5cm,angle=0]{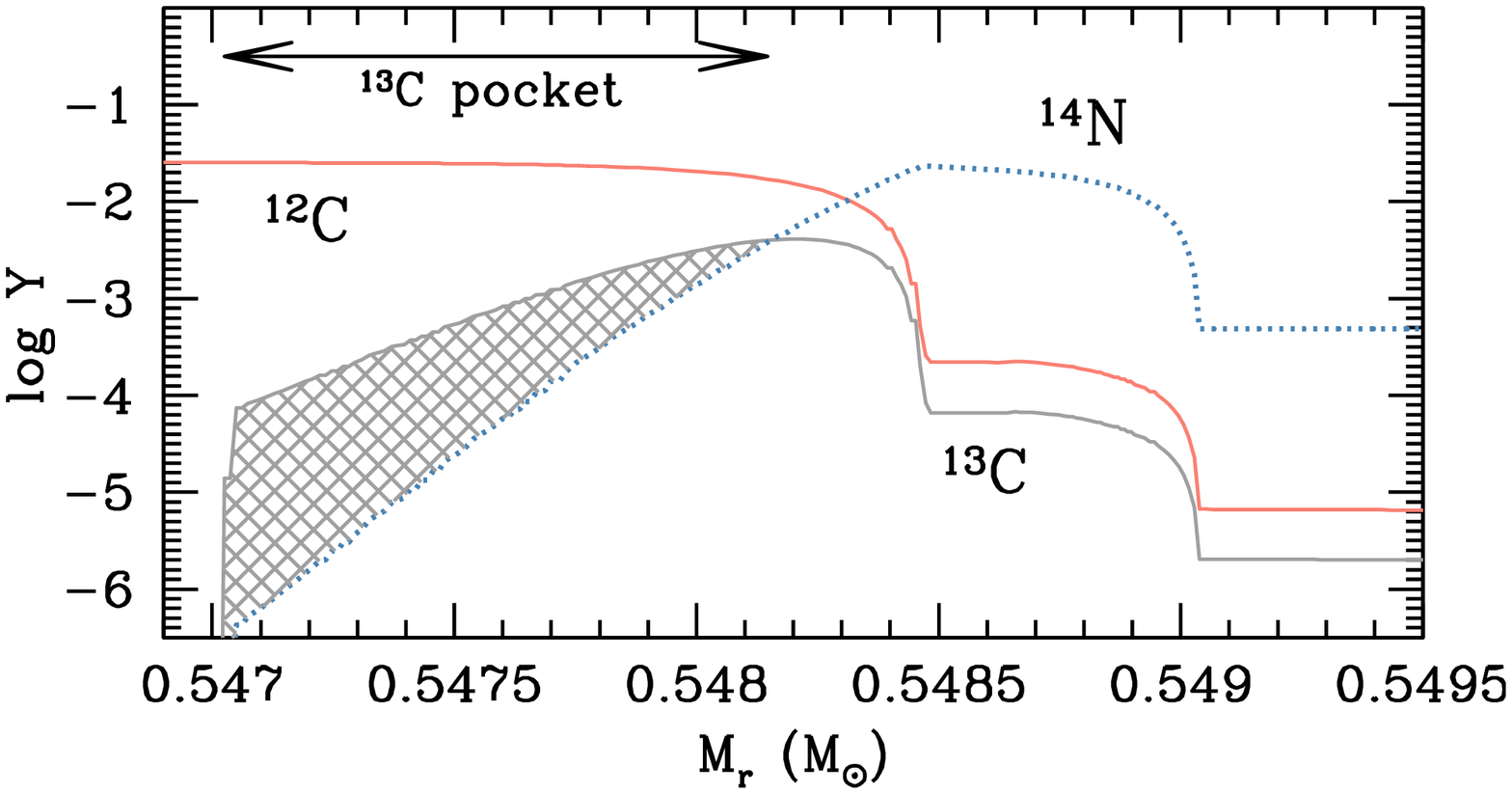}
 \caption{Profile of selected isotopes derived 
using the standard exponential 
mixing function as applied to the 1.25\,M$_\odot$ \emph{Z} = 0.01 model 
at the deepest extent of the first TDU episode. 
Top panel: the proton profile just after insertion of the PMZ, 
the shaded region represents the convective envelope. 
Bottom panel: the $^{13}$C and $^{14}$N pockets 
formed after all protons have burnt. 
The cross hatched 
region represents $M_{\rm tot}{\rm (^{\rm 13}C_{\rm eff})}$ and the double arrow indicates
$M_{\rm pocket}$, as defined in the text.}
 \label{fig:profiles}
 \end{figure}

Clearly, at the top of the PMZ ($m_{\rm scaled}=1$) $10^{f}=1$ and 
at the base ($m_{\rm scaled}=0$) $10^{f}=10^{-4}$. Below $m_{\rm scaled}=0$ we include 
no mixing. Finally, we determine the mixed abundances 
of all species $X_{\rm i}$ in the PMZ using mass conservation:

 \begin{equation}
 X_{\rm i}(m) = 10^{f}\,X_{\rm i}(CE) + (1 - 10^{\emph{f}})\, X_{\rm i}^{\rm intershell}(\emph{m}),
 \end{equation}

where $X_{\rm i}^{\rm intershell}$(\emph{m}) are the original 
abundances in the intershell and $X_{\rm i}(CE)$ the abundances at the base of the convective envelope.
Because $X_{\rm H}^{\rm intershell}$(\emph{m}) = 0, 
for protons the equation becomes simply:

 \begin{equation}
 X_{\rm H}(\emph{m}) = 10^{f}\,X_{\rm H}(CE)
 \end{equation}

which, in logarithmic scale, 
yields a line connecting $X_{\rm H}(CE)$ at \emph{m$_{\rm top}$} and 
$10^{-4}\,X_{\rm H}(CE)$ at \emph{m$_{\rm base}$} 
(top panel of Fig.~\ref{fig:profiles}). 

We use $M_{\rm PMZ} = 2 \times 10^{-3}$\,M$_\odot$ in most of our calculations, which is
within the typical range required to match the 
observation that $s$-process enriched AGB stars of metallicity around solar 
show Ba abundances on average roughly ten times higher than solar 
\citep{smith90a,busso95,lambert95,busso01}.
Models where this parameter has been modified can be found, e.g., in \citet{karakas16}.
In the present paper, 
we will discuss in detail the differences and the connections between varying $M_{\rm PMZ}$ and 
varying the mixing profile, $f$. 

In Fig.~\ref{fig:profiles} we show the standard PMZ included at the deepest 
extent of the first TDU of the 
1.25\,M$_\odot$ \emph{Z} = 0.01 model and the resultant $^{13}$C pocket.
This case is used as an illustrative example in the rest of this 
section and it is typical. 
The figure also illustrates our  
definition of the mass width of $^{13}$C pocket ($M_{\rm pocket}$), 
as the mass width within the PMZ where 
the mole fraction of 
$^{13}$C is greater than the mole fraction of $^{14}$N, and 
the total effective mass of $^{13}$C in the pocket, $M_{\rm tot}{\rm (^{\rm 13}C_{\rm eff})}$. 
The effective mass fraction of $^{13}$C 
at each given mass point is defined following \citet{cristallo09} 
as $X^{13}\rm{C}_{\rm{eff}}$ = $X_{13}$ $-$ 
$X_{14}$\,$\displaystyle\frac{13}{14}$, where $X_{13}$ and $X_{14}$ are the mass fractions 
of $^{13}$C and $^{14}$N, respectively. This definition is required because, as mentioned above, 
\iso{14}N is a 
strong neutron poison via the \iso{14}N(n,p)\iso{14}C reaction.
Integrating X$^{13}\rm{C}_{\rm{eff}}$ over $M_{\rm pocket}$ we obtain
$M_{\rm tot}{\rm (^{\rm 13}C_{\rm eff})}$.

\begin{figure*}
   \begin{center}
 \includegraphics[height=8.5cm,angle=-90]{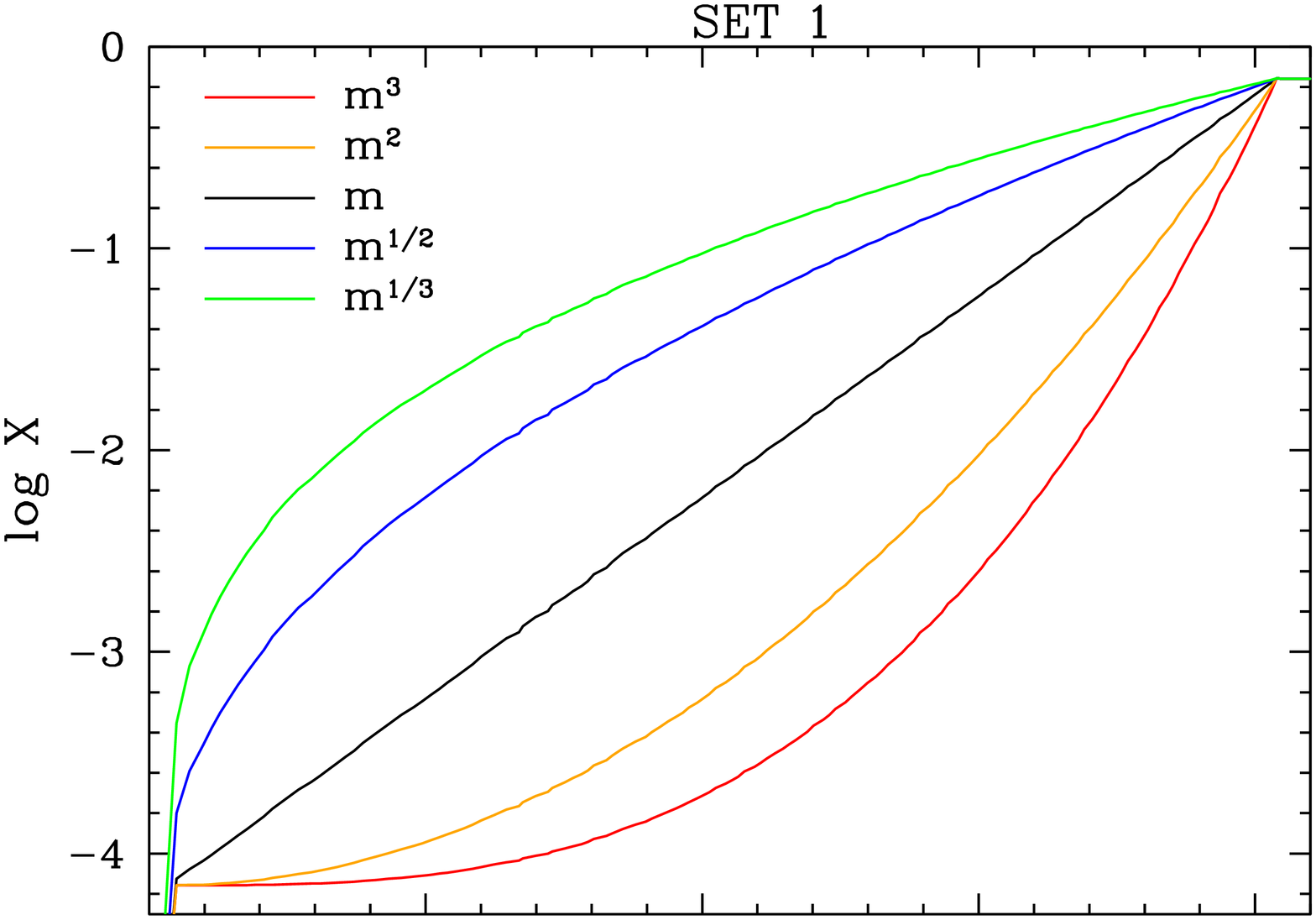} 
 \includegraphics[height=8.5cm,angle=-90]{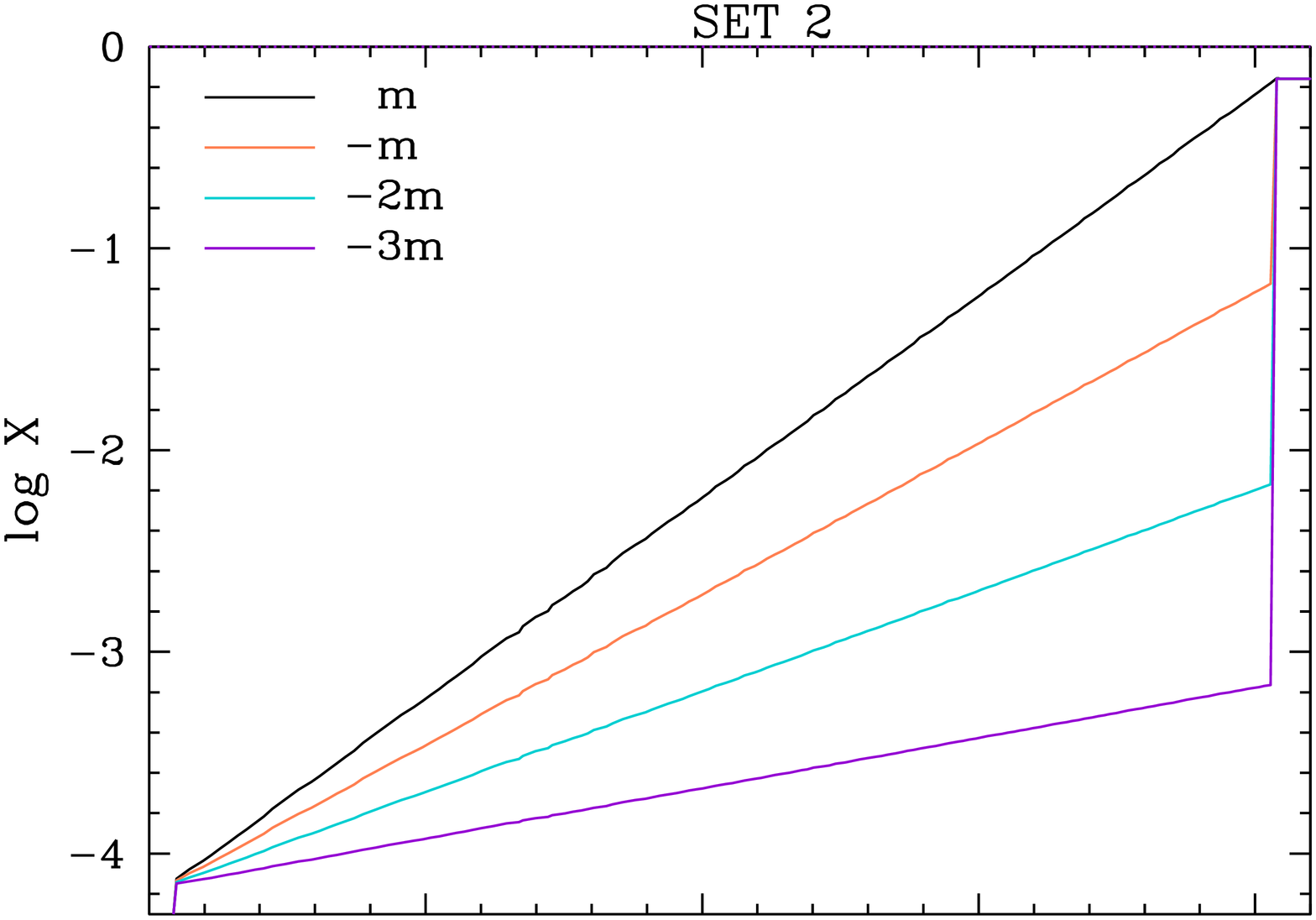}\\
 \includegraphics[height=8.5cm,angle=-90]{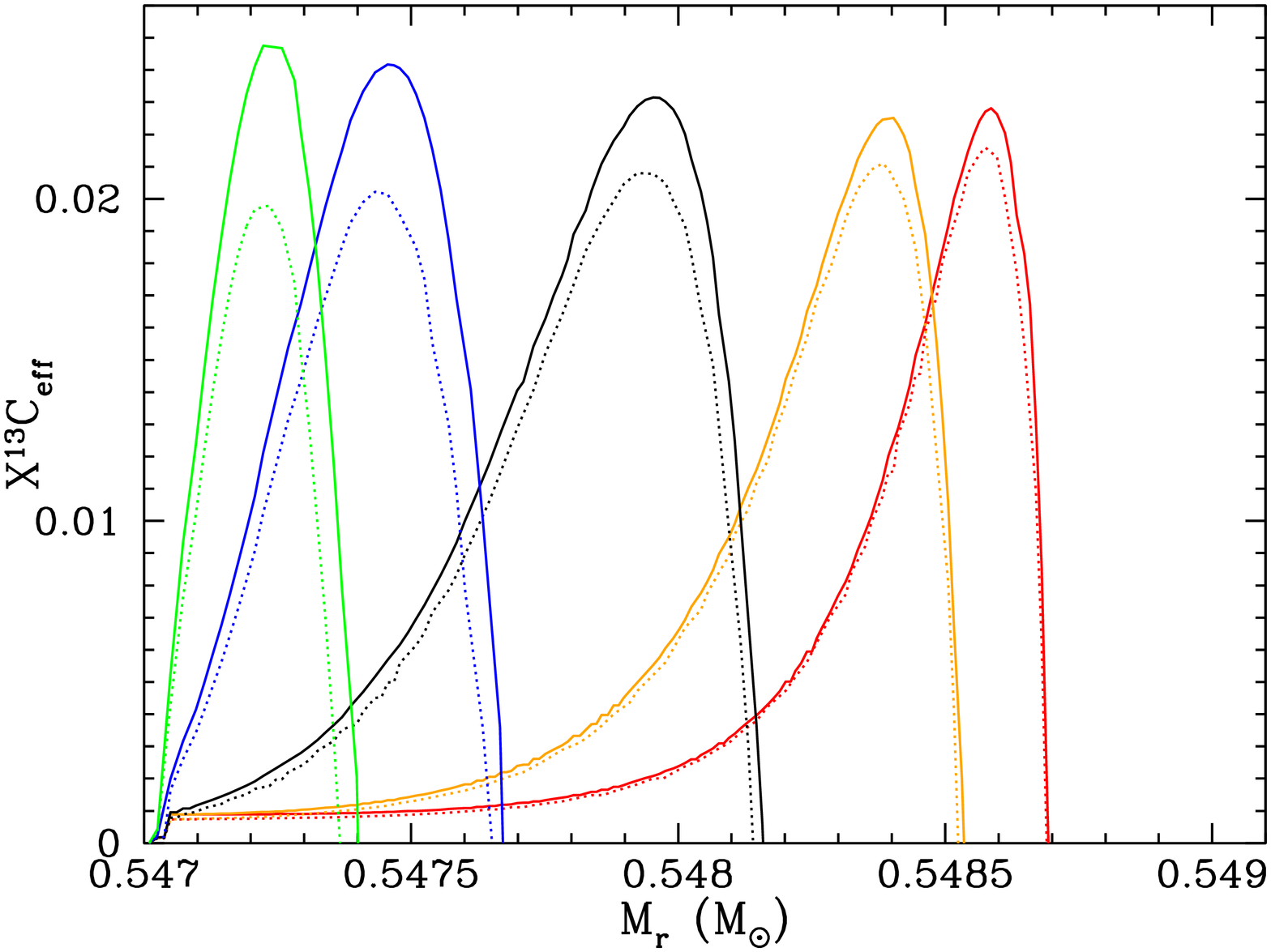} 
 \includegraphics[height=8.5cm,angle=-90]{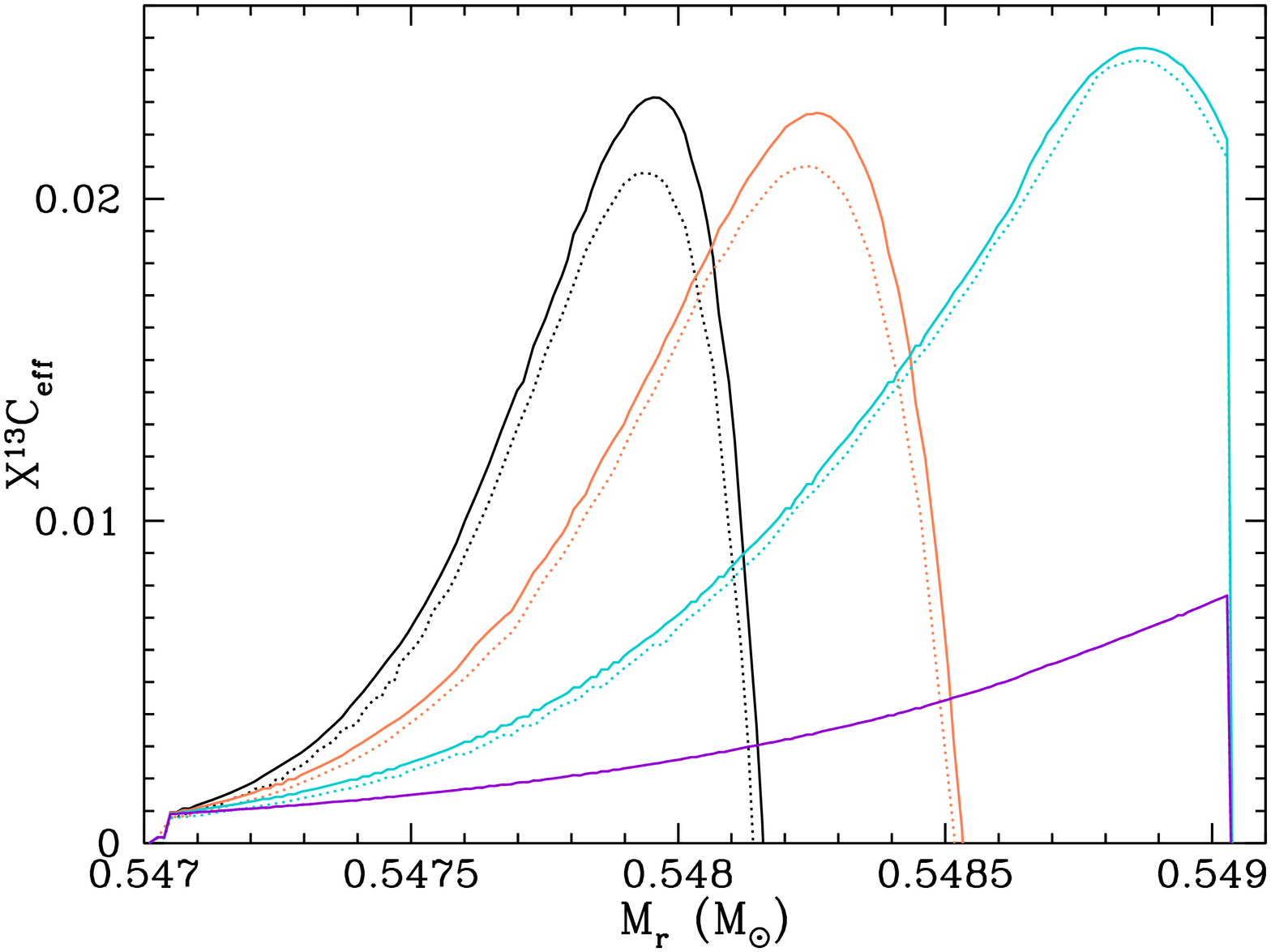}\\
   \end{center}
   \caption{Proton profiles (top panels) and resultant \iso{13}C mass fractions (bottom panels) 
obtained by introducing different shapes of the mixing function of the PMZ 
at the deepest extent of the first TDU episode of the 
1.25 \msun\ Z=0.01 model. Note that the peak in the \iso{13}C mass fraction moves in mass for the 
different cases because it is always located where
log $X_{\rm p} = -2.5$. Above this value of log $X_{\rm p}$, \iso{13}C in converted into \iso{14}N.
The dotted lines in the bottom panels show the 
amount of \iso{13}C remaining at the onset of the next
TP, and subsequently ingested in the TP, a specific feature of the 1.25 \msun\ Z=0.01 model (see text 
for details).}
\label{fig:profiles2}
\end{figure*}

We examined two types of variations from the standard profile: the shape of the mixing profile 
(Set 1) and the introduction of a discontinuity at the top of the PMZ (Set 2). 
In Set 1 we consider the different shaped profiles shown 
in the left panels of Fig.~\ref{fig:profiles2}.
They take the following forms (as compared to Eq.~3) and
we will refer to them hereafter 
using the shorts labels indicated in brackets:

 \begin{equation}
 f = 4\,m^{3}_{\rm scaled} - 4, ~~~~~~~~~~~~~~~~~~~~~~~~(m^{3})
 \label{pp_equations}
 \end{equation}
 \begin{equation}
 f = 4\,m^{2}_{\rm scaled} - 4, \nonumber~~~~~~~~~~~~~~~~~~~~~~~~(m^{2})
 \end{equation}
 \begin{equation}
 f = 4\,m_{\rm scaled} - 4, \nonumber~~~~~~~~~~~~~~~~~~~~~~~~(m, {\rm i.e., standard})
 \end{equation}
 \begin{equation}
 f = 4\,m^{1/2}_{\rm scaled} - 4, \nonumber~~~~~~~~~~~~~~~~~~~~~~~(m^{1/2})
 \end{equation}
 \begin{equation}
 f = 4\,m^{1/3}_{\rm scaled} - 4. \nonumber~~~~~~~~~~~~~~~~~~~~~~~(m^{1/3})
 \end{equation}

Table~\ref{table:c13pocketprop} shows $M_{\rm pocket}$, also in form of  
percentage of the total mass of the PMZ, 
and $M_{\rm tot}{\rm (^{\rm 13}C_{\rm eff})}$
for the different mixing functions.
In Set 1 $M_{\rm pocket}$ decreases when decreasing the 
exponent on the mass, however, 
$M_{\rm tot}{\rm (^{\rm 13}C_{\rm eff})}$
presents a maximum for the standard case because when the exponent on the mass decreases
the size of the $^{14}$N pocket increases.
The $m^2$ case appears to produce the proton 
profile most similar to those obtained
by \citet{cristallo09} for their most efficient choice of $\beta=0.1$ 
(see, e.g., their Fig.~2). 
For their 2 \msun\ model with $Z=0.0138$, 
\citet{cristallo09} finds $M_{\rm PMZ} \simeq 10^{-3}$, 
$M_{\rm pocket}$ = 5.54 $\times 10^{-4}$ \msun\ and 
$M_{\rm tot}{\rm (^{\rm 13}C_{\rm eff})}$ = 7.38 $\times 10^{-6}$ \msun. 
These number are close to those obtained 
with our definition of the mixing function for Set 1, however, it should be kept in mind 
that the values calculated by 
\citet{cristallo09} change for different TDU episodes as the physical properties at the base 
of the convective envelope change and feedback onto the features of the PMZ (see, e.g., their 
Fig.~8). This effect is not considered in our parameteric models, 
where the same PMZ is inserted at each TDU. Interestingly, the fact that the final $s$-process results 
are very similar \citep[as discussed at 
length by, e.g.,][]{karakas16} indicates that this structure feedback 
is second-order effects. 
 
In Set 2, we examine models where a discontinuity is included at the top of the PMZ as shown 
in the right panels of Fig.~\ref{fig:profiles2},
and with features listed in Table~\ref{table:c13pocketprop}.
We construct these profiles as follows (with short notation in brackets): 

\begin{equation}
\emph{f} = 3\,m_{\rm scaled}-4,         ~~~~~~~~~~~~~~~~~~~~~~~~ (-m)
\label{eq:lowpro}
\end{equation}
\begin{equation}
\emph{f} = 2\,m_{\rm scaled}-4, \nonumber~~~~~~~~~~~~~~~~~~~~~~~~ (-2m)
\end{equation}
\begin{equation}
\emph{f} = m_{\rm scaled}-4, \nonumber~~~~~~~~~~~~~~~~~~~~~~~~~ (-3m)
\end{equation}

Among these profiles, the $-3m$ profile is very similar to that obtained by mixing due to 
magnetic fields \citep[see Fig.~2 of][]{trippella16}
and the $-m$ case produces a $^{\rm 13}{\rm C}_{\rm eff}$ profile which is the closest to 
the TAIL case presented by \citet{cristallo15b} (see their Figure~7).

In general, in the Set 2 profiles fewer protons are mixed than in the Set 1 profiles, which results in 
a lower amount of $^{14}$N and a larger $M_{\rm pocket}$. A higher $M_{\rm tot}{\rm (^{\rm 13}C_{\rm eff})}$
is also obtained, 
except for the $-3m$ case, where the lowest number of 
protons are mixed.

\begin{table}
\centering
\caption{Properties of the $^{13}$C pockets for the different mixing profiles. Masses are all in \msun.}
\setlength{\tabcolsep}{2.5pt}
\begin{tabular}{lccc}
\hline 
Profile & $M_{\rm pocket}$ & \% of $M_{\rm PMZ}$ & $M_{\rm tot}{\rm (^{\rm 13}C_{\rm eff})}$ \\
 & (10$^{-3}$) & & (10$^{-6}$) \\
\hline 
 \multicolumn{4}{c}{\bf Set 1} \\ 
\hline 
$m^{3}$ & 1.70 & 85\% & 8.58  \\
$m^{2}$  &  1.53 & 77\% & 9.92 \\
$m$     & 1.16 & 58\% & 11.6 \\
$m^{1/2}$ & 0.67 & 33\% & 9.31 \\
$m^{1/3}$ & 0.40 & 20\% & 5.97 \\ 
\hline 
\multicolumn{4}{c}{\bf Set 2} \\ 
\hline 
$m$     & 1.16 & 58\% & 11.6 \\
$-m$ & 1.53 & 76\% & 15.2 \\
$-2m$ & 1.99 & 99\% & 20.5 \\
$-3m$ & 2.00 & 100\% & 6.33 \\
\hline 
\end{tabular}
\label{table:c13pocketprop}
\end{table}

\section{Results}

\subsection{Neutron density and neutron exposure during the AGB phase}

The initial stellar mass and metallicity determine the main evolutionary features during the 
AGB phase, and these in turn determine which neutron source is activated, how the \iso{13}C burns, 
the resulting neutron density N$_{n}$, 
and total number of free neutrons. The latter is usually measured via the neutron exposure $\tau$, 
defined as the time-integrated neutron flux $\int {\rm N}_{n}\,\upsilon_{T} \,dt$, where  
$\upsilon_{T}$ is the thermal velocity.
 
A summary of the sources of the neutron fluxes present during the AGB phase in the different models 
is presented in Table~\ref{table:models_burn} \citep[see also][for a detailed description of the different
regimes]{lugaro12}. In 
relation to the \iso{13}C pocket we find in our models both the  
uncommon case, specific 
to stars of low mass and high metallicity where some \iso{13}C is left in the pocket and burns while 
ingested in the next TP \citep[\iso{13}C $ingestion$][]{cristallo09}, as originally proposed as the standard 
scenario \citep{iben82} and 
the common case where all the \iso{13}C in the 
pocket burns in radiative conditions before the onset of the next TP \citep[\iso{13}C $radiative$, as 
discovered by][]{straniero95}.
At low mass and metallicity, we also find a few instances 
of proton ingestion episodes directly inside the TPs. These protons 
produce extra \iso{13}C and neutrons. For stars of mass higher than roughly 3 \msun, 
when the maximum temperature in the 
intershell reaches above 300 MK (see Table~\ref{table:models}), neutrons are also released 
by the \iso{22}Ne 
neutron source inside the TPs (\iso{22}Ne burning). 

\begin{table}
\centering
\caption{
Summary of the sources of the neutron fluxes experienced by the different stellar models.}
\setlength{\tabcolsep}{2.5pt}
\begin{tabular}{llllll}
\hline 
Mass & \emph{Z} & $^{13}$C & $^{13}$C & proton & $^{22}$Ne\\
& & radiative & ingestion & ingestion &  burning \\
\hline 
\centering
1.25 & 0.01 &  & all PMZs & & \\
1.8 & 0.01 & late PMZs & early PMZs & & \\
3.0 & 0.01 & all PMZs &  &  & late TPs \\
3.0 & 0.02 & all PMZs &  &  & late TPs \\
4.0 & 0.02 & all PMZs &  & & late TPs \\ 
1.5 & 0.0001 & late PMZs & early PMZs & early TPs & \\\hline 
\end{tabular}
\label{table:models_burn}
\end{table}

\begin{figure}
  \begin{center}
\includegraphics[width=6.5cm,angle=-90]{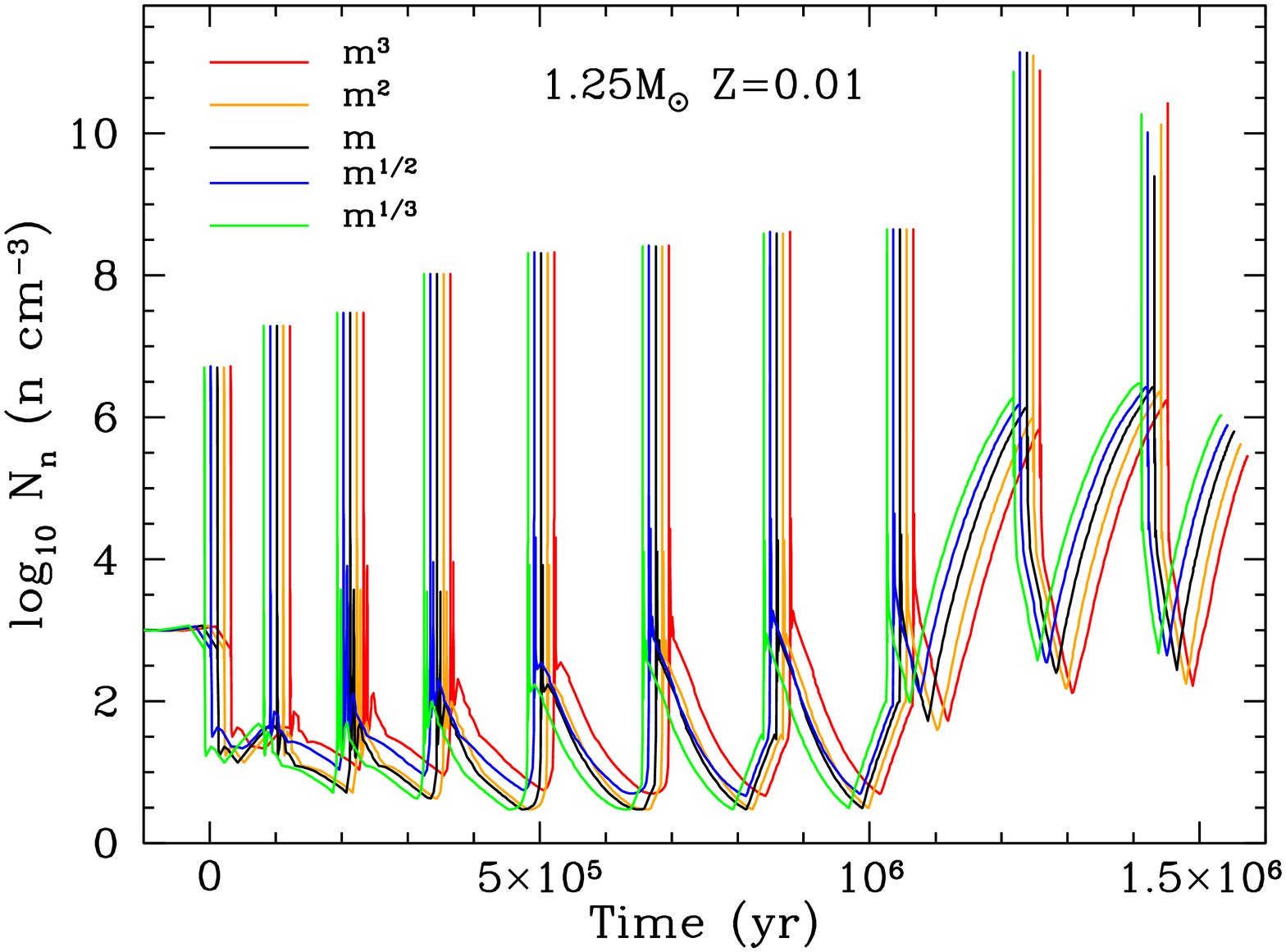}
 \includegraphics[width=6.5cm,angle=-90]{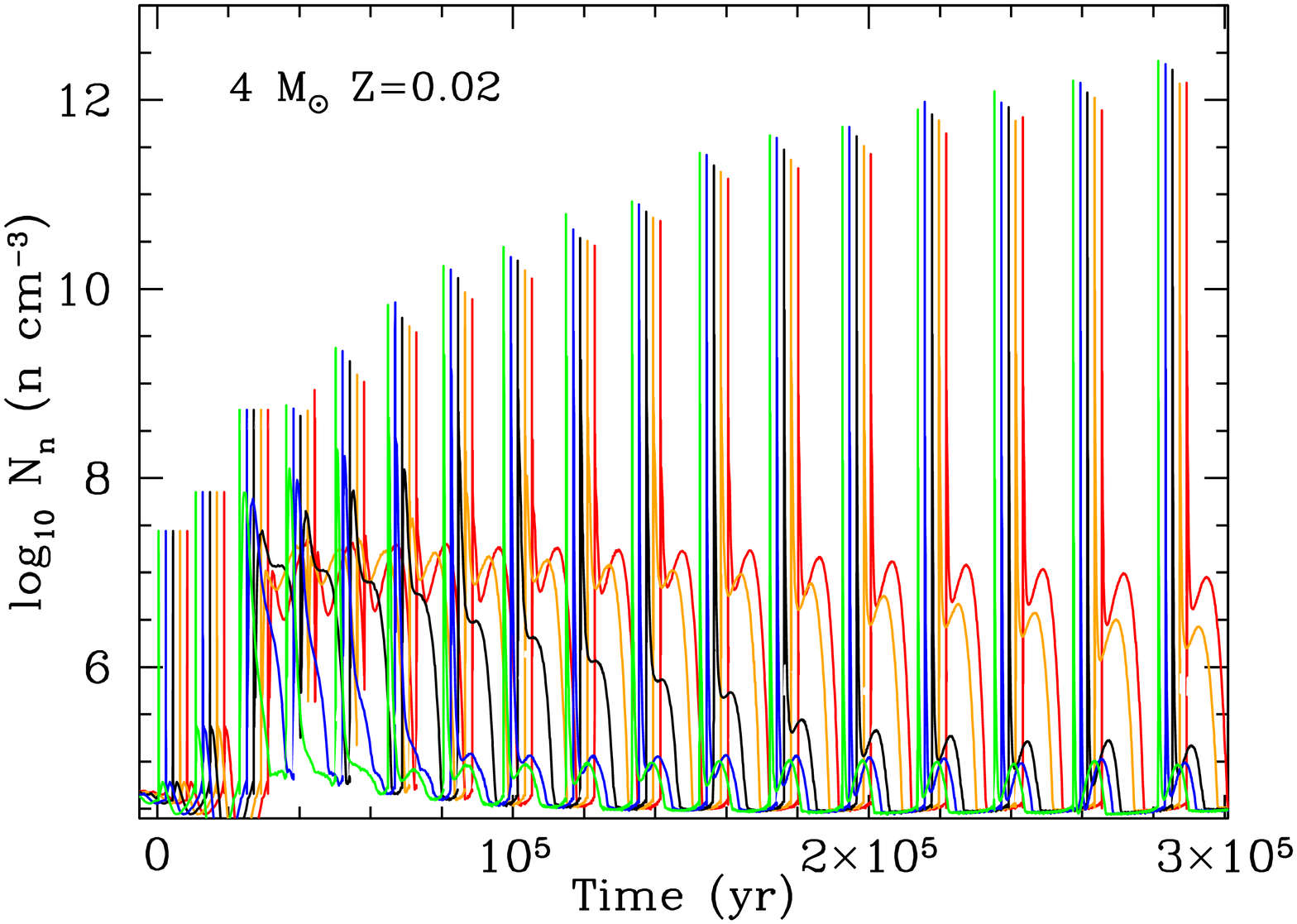}
\includegraphics[width=6.5cm,angle=-90]{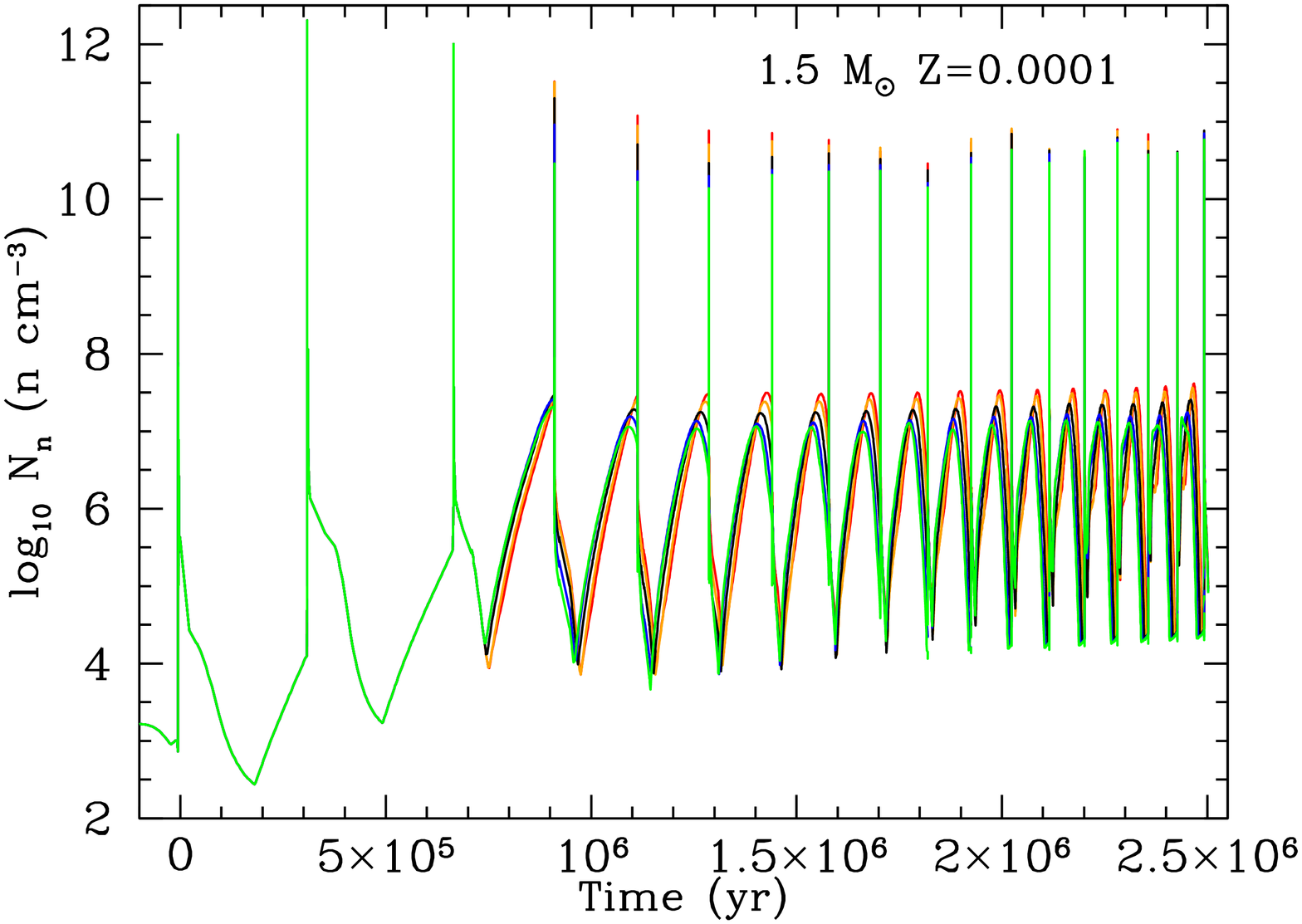}
  \end{center}
  \caption{The maximum neutron density within the star 
as function of time during the thermally pulsing AGB phase of the 1.25\,M$_\odot$ 
\emph{Z} = 0.01 (top panel), 4\,M$_\odot$ \emph{Z}= 0.02 (middle panel) and  
1.5\,M$_\odot$ \emph{Z}= 0.0001 (bottom panel) models for the different cases of Set 1. 
The zero time is set at the time of the 1st TP for the $m^{1/3}$ case. 
To be able to distinguish the different cases, 
the other lines are all slighly offset in time. Spikes in the N$_{n}$ identify the TPs.
Note the 
difference in the duration of the TP-AGB phase between the models, which increases with decreasing  
mass and/or metallicity.}
\label{fig:neutrondensity}
\end{figure}

In Fig.~\ref{fig:neutrondensity} we show, for three representative stellar models
calculated using Set 1, the evolution as a function of time 
of the maximum N$_{n}$ within the whole star, which is the best proxy to identify 
the different significant neutron flux events along the AGB evolution.
Table~\ref{table:tau} lists the values of the neutron exposure calculated 
during selected interpulse periods 
using the maximum neutron density shown in Fig.~\ref{fig:neutrondensity}. 
We selected early and late \iso{13}C pockets during the evolution 
to show how the neutron flux changes due to 
both structural and feedback effects along the evolution.
We discuss each example star in a separate subsection.

\begin{table}
\centering
\caption{The neutron exposure $\tau$ (in mbarn$^{-1}$) during the interpulse periods for all the models
($M$,$Z$ stands for the mass in \msun\ and the metallicity of the star).
For the models that experience more than five TDUs episodes (hence PMZs) values are given for 
two different PMZs, selected as representative of an early and a late time during the 
evolution.} 
\setlength{\tabcolsep}{2.5pt}
\begin{tabular}{lccccccccc}
\hline
$M$,$Z$ & PMZ & $m^{3}$ & $m^{2}$ & $m$ & $m^{1/2}$ & $m^{1/3}$ & $-m$ & $-2m$ & $-3m$ \\
\hline
1.25,0.01  & 1  & 0.04 & 0.06 & 0.10 & 0.13 & 0.16 & 0.06 & 0.04 & 0.01 \\
\hline
1.8,0.01   & 5  & 0.59 & 0.59 & 0.58 & 0.56 & 0.53 & 0.68 & 0.70 & 0.30 \\
\hline
3.0,0.01   & 1  & 0.63 & 0.67 & 0.71 & 0.72 & 0.68 & 0.80 & 0.77 & 0.36 \\
3.0,0.01   & 16 & 0.66 & 0.68 & 0.74 & 0.67 & 0.57 & 0.78 & 0.80 & 0.40 \\
\hline
3.0,0.02   & 1  & 0.47 & 0.48 & 0.50 & 0.49 & 0.46 & 0.59 & 0.50 & 0.22\\
3.0,0.02   & 15 & 0.45 & 0.44 & 0.42 & 0.38 & 0.34 & 0.54 & 0.60 & 0.25 \\
\hline
4.0,0.02   & 1  & 0.54 & 0.57 & 0.58 & 0.54 & 0.48 & 0.74 & 0.70 & 0.35 \\
4.0,0.02   & 14 & 0.62 & 0.69 & 0.45 & 0.07 & 0.04 & 0.97 & 1.13 & 0.50 \\
\hline
1.5,0.0001 & 1  & 2.35 & 2.78 & 3.64 & 3.95 & 3.34 & 4.03 & 3.45 & 2.03 \\
1.5,0.0001 & 15 & 2.58 & 2.31 & 1.87 & 1.42 & 1.23 & 3.38 & 4.25 & 2.80 \\
\hline
\end{tabular}
\label{table:tau}
\end{table}

\subsubsection{The 1.25 \msun\ Z=0.01 model}

The spikes in the neutron density in the top panel of Fig.~\ref{fig:neutrondensity} 
clearly show the 10 TPs experienced by this model.
The N$_{n}$ spikes corresponding to the first 8 TPs represent 
the signature of the convective burning of the \iso{13}C from the 
H-burning ashes, resulting in 
a low neutron density (10$^{7-8}$\,cm$^{-3}$) and insignificant neutron 
exposure. After the 8th TP, a TDU episode occurs and the PMZ is inserted. 
At this point the radiative burning of \iso{13}C is  
noticeable as a 
smooth increase in N$_{n}$ during the interpulse periods. However,
due to temperatures in the pocket being below 80 MK, 
only a small fraction of  
\iso{13}C burns before the onset of the following TP and 
a significant amount of \iso{13}C is still left when the 
next TP occurs. As shown in the bottom panels of Fig.~\ref{fig:profiles2} 
this fraction depends on the PMZ case used, with a 
higher fraction left for the case $m^3$, which has the both the lowest local abundance of \iso{13}C 
- the rate of the 
\iso{13}C($\alpha$,n)\iso{16}O reaction is proportional to the number of \iso{13}C nuclei -
and the maximum value of \iso{13}C located at the highest mass, where the temperature is lower. 

The difference in the \iso{13}C fraction left to be ingested 
between the different Set 1 cases is reflected in the values of 
the neutron exposures during the interpulse periods, which increase from 0.04 
mbarn$^{-1}$ for case $m^3$ to 0.16 mbarn$^{-1}$ for case $m^{1/3}$ 
(Table~\ref{table:tau}).
The ingestion of the \iso{13}C (and \iso{14}N) pocket in the following TPs 
produces a higher spike in N$_{n}$ during the last 2 TPs than in the previous TPs, 
up to roughly 10$^{11}$\,cm$^{-3}$, depending on the TP and the PMZ case considered -- i.e., 
depending on the 
interplay of the \iso{13}C($\alpha$,n)\iso{16}O, \iso{14}N($\alpha$,$\gamma$)\iso{18}F,  
\iso{14}N(n,p)\iso{14}C reactions, and the mixing timescale as different 
\iso{13}C and \iso{14}N abundances are ingested in the TPs.

Also the 1.8\,M$_\odot$ \emph{Z} = 0.01 model experiences a few \iso{13}C ingestions during the early TPs, 
however, their overall 
effect is much smaller than in the case of the 1.25\,M$_\odot$ \emph{Z} = 0.01. The 
neutron exposure in the \iso{13}C pocket during the interpulse periods for this model is
$\sim$ 0.60 mbarn$^{-1}$, very close to that experienced by the 3\,M$_\odot$ \emph{Z} = 0.01 model 
($\sim$ 0.70 mbarn$^{-1}$), which does not suffer any \iso{13}C ingestion.
In these cases, where \iso{13}C radiative burning is the main mode of neutron production,
the neutron exposure in the \iso{13}C
pocket does not change significantly with the mixing profile because it is controlled by the maximum value
of $X^{13}C_{\rm eff}$, which is relatively constant (Fig.~\ref{fig:profiles2}), although feedback effects play a role 
in some of the models, as discussed below. 

\subsubsection{The 4 \msun\ Z=0.02 model}

The middle panel of Fig.~\ref{fig:neutrondensity} shows the evolution of the 
maximum neutron density during the 17 
TPs experienced by this model for the different cases of Set 1. As the TP temperature increases during 
the evolution, the effect of the $^{22}$Ne neutron source becomes more visible in the later TPs, where N$_{n}$ 
reaches above 10$^{12}$\,cm$^{-3}$. The N$_{n}$ in the TPs 
increases from the $m^3$ to the $m^{1/3}$ 
cases because the \iso{14}N pocket becomes larger and contributes to the abundance 
of $^{22}$Ne in the TPs, increasing the rate of the \iso{22}Ne($\alpha$,n)\iso{25}Mg reaction.

\begin{figure}
  \begin{center}
\includegraphics[width=8.5cm]{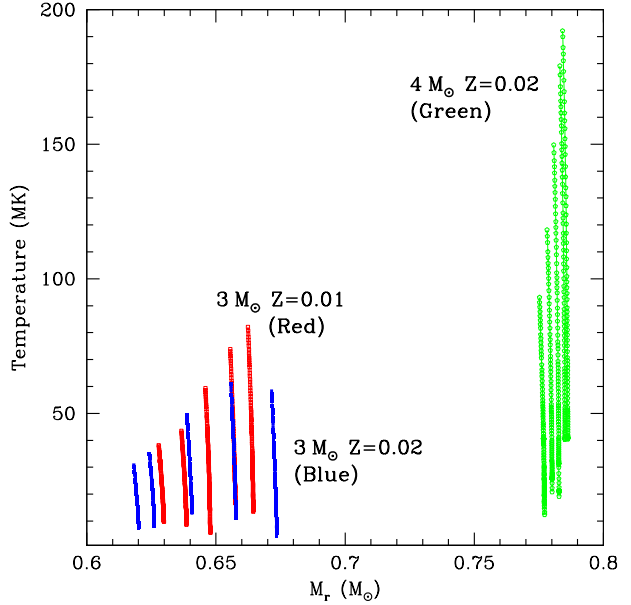}
  \end{center}
  \caption{The temperature as a function of mass in the PMZ at the time when the PMZ is inserted 
for the 3\,M$_\odot$ models of \emph{Z} = 0.01 (red) and 0.02 (blue), and the 
4\,M$_\odot$ \emph{Z} = 0.02 model (green). 
Each line represents the temperature within the PMZ for some representative  
TDU episodes along the evolution: for the 3\,M$_\odot$ model of \emph{Z} = 0.02 models, the 
1st, 2nd, 5th, 10th and 16th TDU, for the 3\,M$_\odot$ \emph{Z} = 0.01 and 4\,M$_\odot$ \emph{Z} = 0.02 
models, the 1st, 3rd, 
6th, 10th, and 13th TDU, with the TDU number increasing from left to right, i.e., 
in order of increasing the location in mass coordinate $M_{\rm r}$.}
\label{fig:temperature}
\end{figure}

Distinct differences are present in the N$_{n}$ produced in the \iso{13}C pocket during the 
interpulse periods for the different Set 1 cases: the two most extreme cases $m^{3}$ and $m^{1/3}$ show neutron 
densities during the slow burning phase within the 
interpulse periods typically as high as $10^7$ and as low as $10^5$\,cm$^{-3}$, respectively. 
The reason is a combination of two factors: (1) 
depending on the choice of the PMZ profile the maximum \iso{13}C abundance is located closer to (e.g., case
$m^{3}$) or further from (e.g., case $m^{1/3}$) 
the bottom of the envelope (Fig.~\ref{fig:profiles2}) and (2) the temperature in the region where the PMZ 
is inserted is a steep function of the location in mass and is much higher in the 4 \msun\ model than in the lower 
mass models (Fig.~\ref{fig:temperature}). It follows that in the $m^{1/2}$ and $m^{1/3}$ cases, which 
form the $^{13}$C 
pocket deeper in the star than the other cases, $^{13}$C is immediately exposed to the temperatures at which 
the \iso{13}C($\alpha$,n)\iso{16}O reaction is activated. A spike in the neutron density occurs already 
on a very short timescale in correspondence to the introduction of the PMZ (in the figure, the spike 
is very close to the spike corresponding to the previous TP produced by the \iso{22}Ne neutron source). 
This effect becomes even more 
pronounced as the star evolves and the temperature increases (Fig.~\ref{fig:temperature}). In the late TPs 
for the $m^{1/2}$ and $m^{1/3}$ cases, the temperature at which the PMZ is inserted is so high
that the CN cycle is established, \iso{13}C is quickly transmuted into 
\iso{14}N, and the amount of \iso{13}C becomes smaller (e.g., $M_{\rm pocket}$ 
decreases by a factor of five or so). 
In this situation the neutron flux is 
effectively inhibited, and the neutron exposure becomes lower than 0.1 mbarn$^{-1}$ for the 
$m^{1/2}$ and $m^{1/3}$ cases (see Table~\ref{table:tau}).
In the $m^2$ and $m^3$ cases, instead, the fewer number of protons mixed do not lead to the same 
very efficient burning
at the time of the formation of the PMZ and a higher neutron flux results during the interpulse periods. 
The $m$ proton profile is an intermediate case, which
also shows the effect of the time evolution of the temperature within each PMZ and
the \iso{13}C pocket. When the PMZ is inserted, the temperature is high enough to activate  
the \iso{13}C($\alpha$,n)\iso{16}O reaction and a spike in the N$_{n}$ is seen. As the 
envelope receedes after the TDU, the temperature decreases, to start increasing again from the start of the 
activation of the H-burning shell. 

Finally, we note that the 4 \msun\ $Z=0.02$ model reaches 40 MK at most at the base of the 
envelope during the latest TDUs, which is only bordeline to trigger the ``hot TDU'' \citep{goriely04}.
In the case of the hot TDU, 
the CN cycle is at work during the mixing responsible for the formation of the PMZ
and the formation of the \iso{13}C pocket is 
strongly inhibited. 
In any case, as 
derived above by comparing, e.g., 
the $m^{1/3}$ to the $m^3$ case in the middle panel of Fig.~\ref{fig:neutrondensity}
and in Table~\ref{table:tau} we find that, strongly depending on the mixing profile, burning processes other 
than the hot TDU are also 
at work during the formation of the PMZ, which can suppress the number of neutrons released in the 
\iso{13}C pocket in the 4 \msun\ model.


\subsubsection{The 1.5 \msun\ Z=0.0001 model}

The bottom panel of Fig.~\ref{fig:neutrondensity} shows the neutron density as a function of time for the 18 TPs 
experienced by the 1.5\,M$_\odot$ \emph{Z} = 0.0001 model. The very low metallicity of this model 
plays an important role in producing three to five times higher neutron exposures than the models of higher 
metallicity (Table~\ref{table:tau}) due to the primary nature of 
the \iso{13}C neutron source. 
The $m^{1/3}$ proton profile produces the largest \iso{14}N pocket, which results in a
strong enhancement of primary \iso{22}Ne in the intershell. The 
\iso{22}Ne(n,$\gamma$)\iso{23}Ne($\beta^{-}$)\iso{23}Na reaction chain steals neutrons from the $s$-process,
while increasing the production of Na (see Table~\ref{table:Na}) with the consequence that the 
neutron exposures decreases during the evolution ($3.34-1.23$ mbarn$^{-1}$).
Conversely, the $m^{3}$ proton profile produces the smallest \iso{14}N pocket so the neutron expsoure 
does not significantly changes during the evolution ($2.35-2.38$ mbarn$^{-1}$). 
It 
is also noticeable in Fig.~\ref{fig:neutrondensity} that the early TPs experience a spike in the 
neutron density, reaching up to $\sim 10^{12}$ cm$^{-3}$ before the PMZ is inserted.
We infer these to be caused by mild proton ingestions. Furthermore, 
the effect of \iso{13}C ingestions are visible  
in the first few TPs after the PMZ is inserted, and explain the fact that the 
neutron exposure during the interpulse period is smaller for the 2$^{\rm nd}$ \iso{13}C pocket than for the 
last (Table~\ref{table:tau}). However, the final $s$-process distribution 
produced by this model is dominated by the effect of the radiative \iso{13}C 
burning, which produces the neutron flux noticeable from the figure in most of the interpulse periods.

\subsection{Abundance results for the elements heavier than Fe}

In Figs.~\ref{fig:results1} and \ref{fig:results2} we present the elemental abundances in the form 
[X/Fe]\footnote{[X/Fe] is defined as
log($N_{\rm X}/N_{\rm Fe})_{\rm star}$ $-$ log($N_{\rm X}/N_{\rm Fe})_{\odot}$, where $N_{\rm X}$ and 
$N_{\rm Fe}$ are the
abundances by number of element X and of Fe, respectively.} resulting at the stellar 
surface at the end of the evolution for all our calculated models. 
We will focus on the $s$-process elements
Sr, Ba, and Pb, as representative of the first, second, and third $s$-process peaks, respectively.
We discuss in our analysis the 
absolute enrichments, i.e., [Sr/Fe], [Ba/Fe], and
[Pb/Fe], and the relative abundance distribution, i.e.,  [Ba/Sr] and [Pb/Ba], describing 
the relative height of the three $s$-process peaks. 
These ratios 
are reported in Table~\ref{table:abund} for all the computed models. 

The initial stellar mass and metallicity play the major role in determining the final $s$-process abundances.
The stellar structure (mainly the 
temperature) is responsible for activating the different types of neutron sources, which result in 
fluxes of neutrons characterised by the different neutron densities and exposures detailed above.
We can derive the main differences between the stellar models in terms of both 
absolute abundances and relative
distribution by comparing the standard $m$ cases shown in the different 
panels of Fig.~\ref{fig:results1}. 
These differences can be summarised as follows: (1) The abundance distribution for the 
1.25\,M$_\odot$ is almost fully determined by the \iso{13}C ingestions, which produce higher 
neutron densities and lower neutron exposures than the radiative \iso{13}C burning occurring in all 
the other models. For this reason the final $s$-process distribution is shifted towards the first 
$s$-process peak at Sr. With respect to the 1.8 and 3 \msun\ models, the higher neutron density
also produces higher [Rb/Zr] ratios, due to the activation of the branching point at \iso{86}Rb.
A further consequence of the \iso{13}C ingestions, together with the fact that this model 
experiences the lowest amount of TDU of all the models (Table~\ref{table:models}), is that the absolute 
abundances are lower by roughly one order of magnitude than in the other models. 
(2) Because the neutron exposure increases
with metallicity, the 1.5\,M$_\odot$ 
$Z=0.0001$ model has the highest abundance of the third peak element Pb. It also has 
the highest absolute abundances by one order of magnitudes. This is mostly due to the normalisation to Fe, 
which is two orders of magnitude lower than in the other models. 
Metallicity driven differences are also noticeable in the relative abundance 
distribution between the $Z=0.01$ and $Z=0.02$ 
models, e.g., the [Ba/Sr] ratio is three times higher in the former than the latter.
(3) Between the two 
$Z=0.02$ models, the 4 \msun\ model shows a marginally higher enrichment relative to the
3 \msun\ model in the second and third 
peaks, by at most 50\% relative to the first peak, due to the higher contribution of the 
\iso{22}Ne neutron source.

Overall, it is the combination of the different types of neutron fluxes, their relative 
contribution, and the efficiency of the TDU that control the final results as a function of the stellar 
mass and metallicity. On top of these, here we aim to identify and discuss   
the effects of changing the features of the PMZ for 
each stellar model.

\subsubsection{Comparison of the results from Set 1}

\begin{figure*}
\centering
\includegraphics[width=6.5cm,angle=-90]{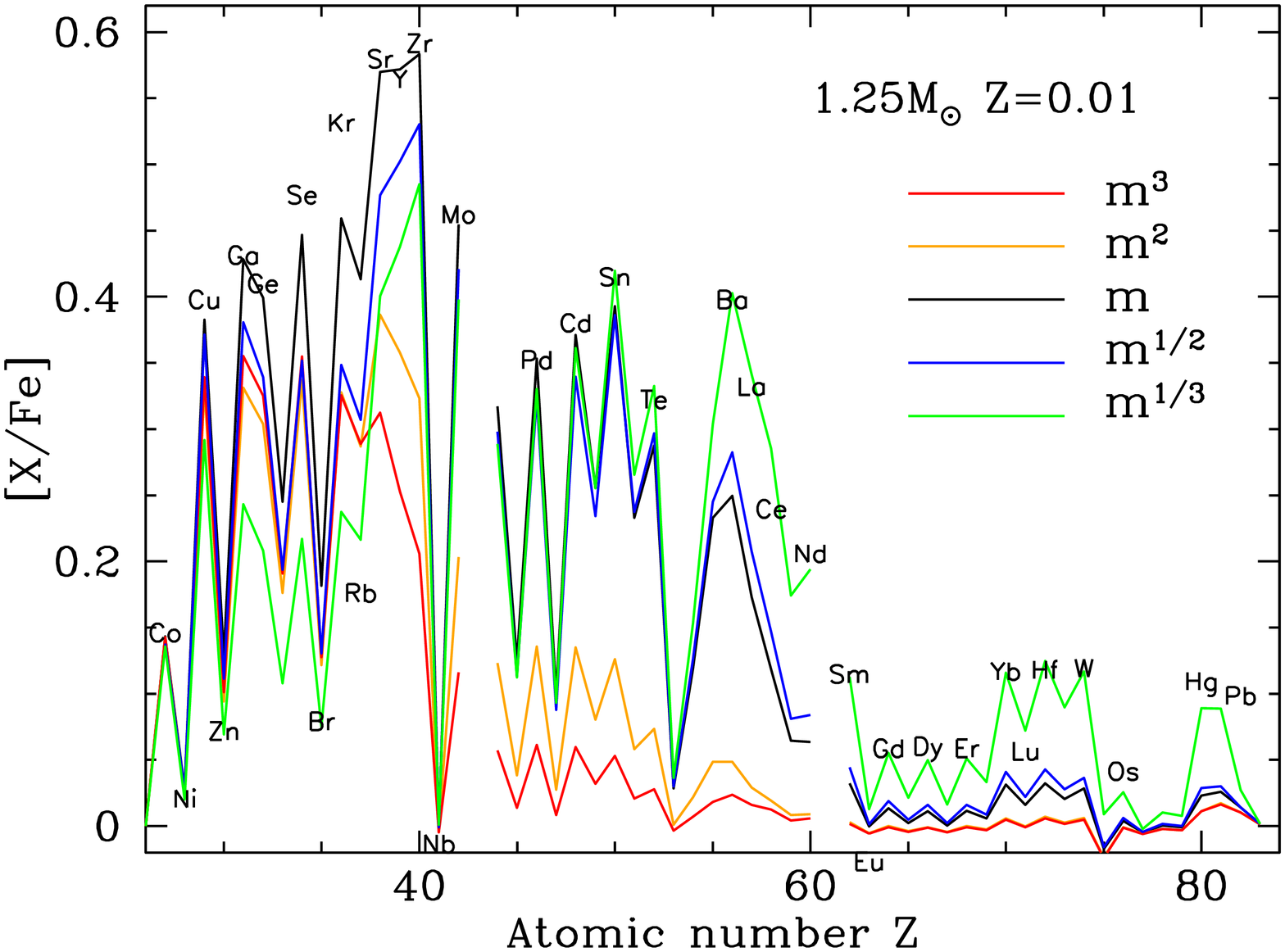} 
\includegraphics[width=6.5cm,angle=-90]{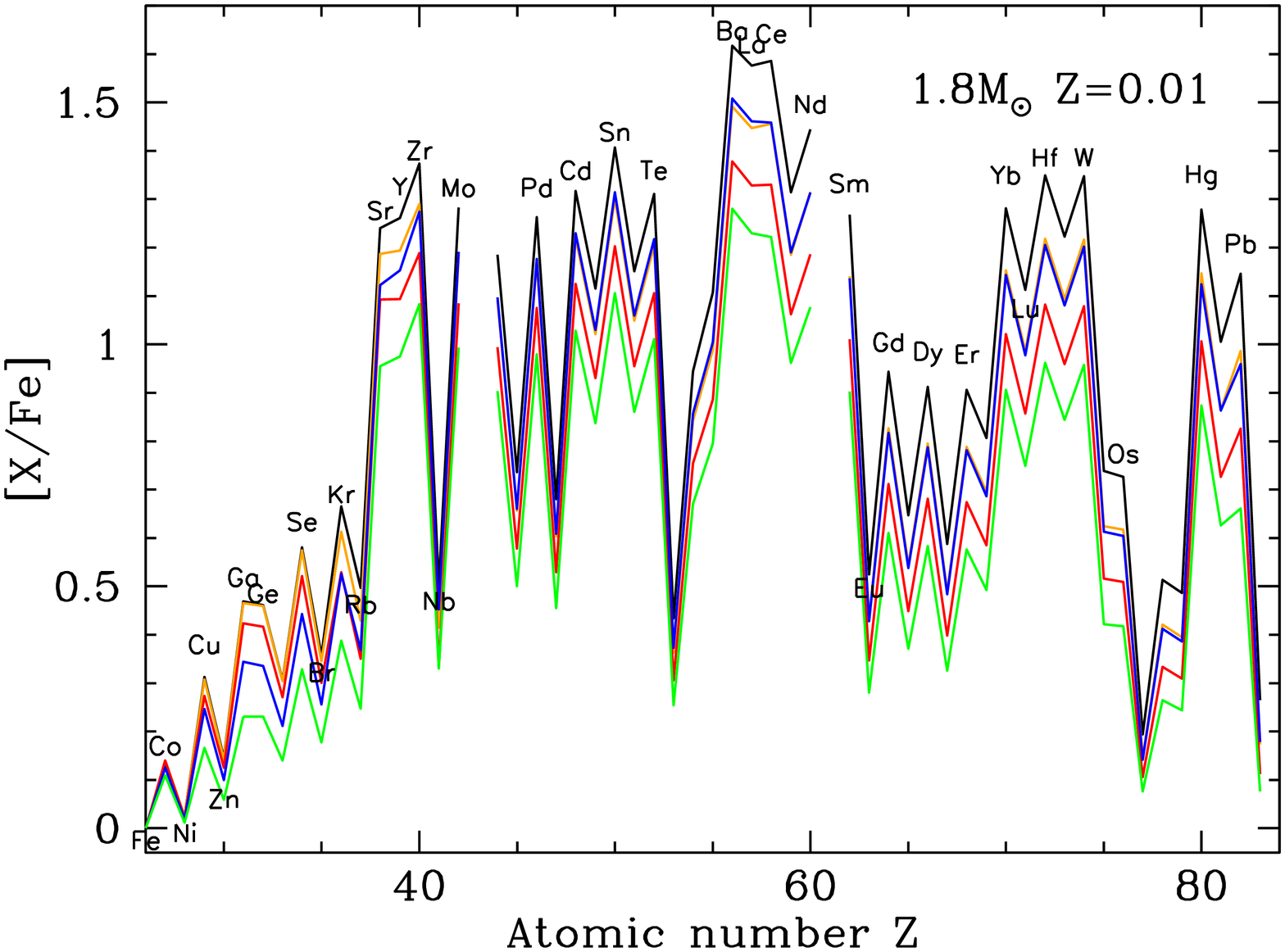} \\
\includegraphics[width=6.5cm,angle=-90]{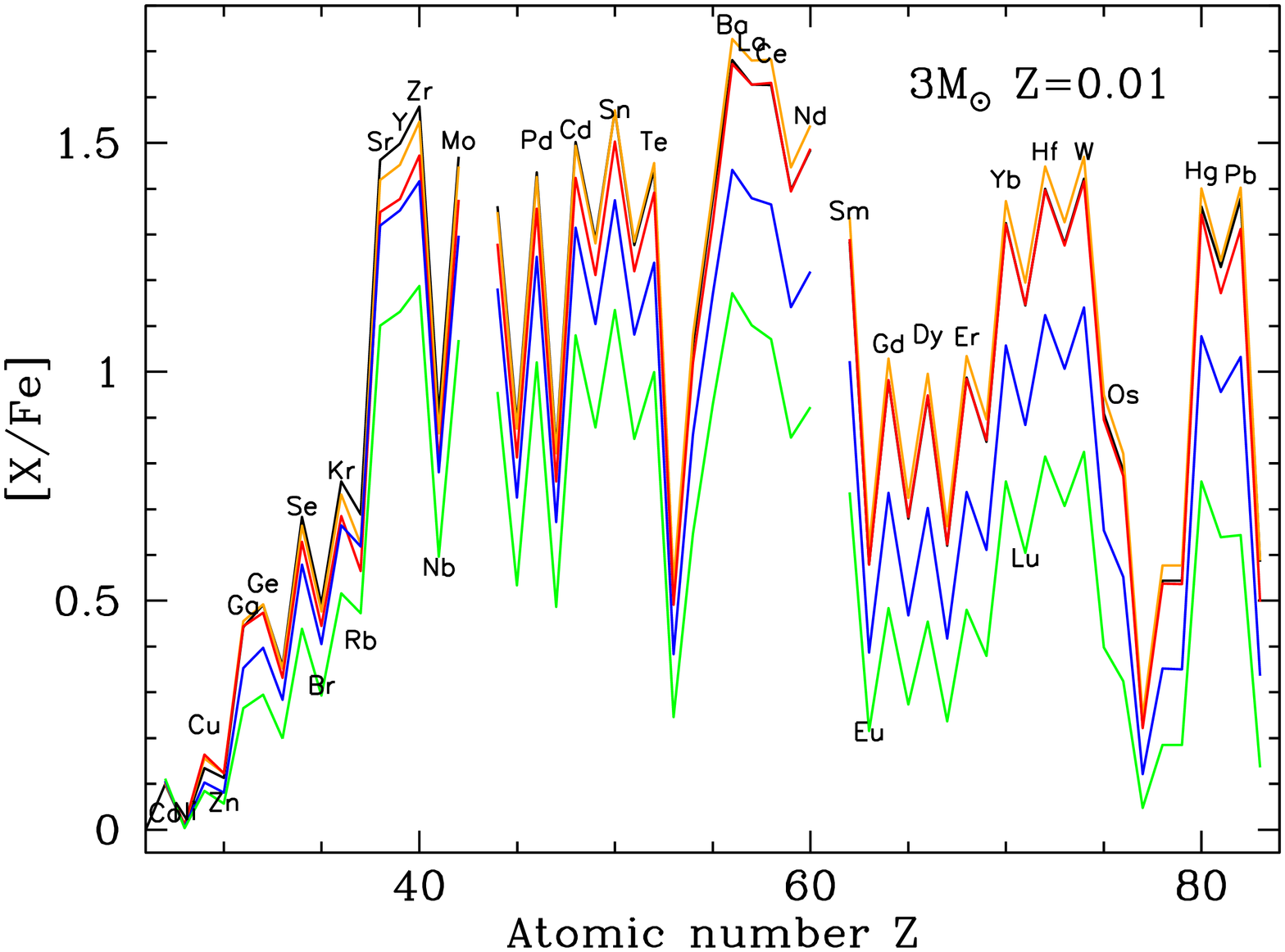} 
\includegraphics[width=6.5cm,angle=-90]{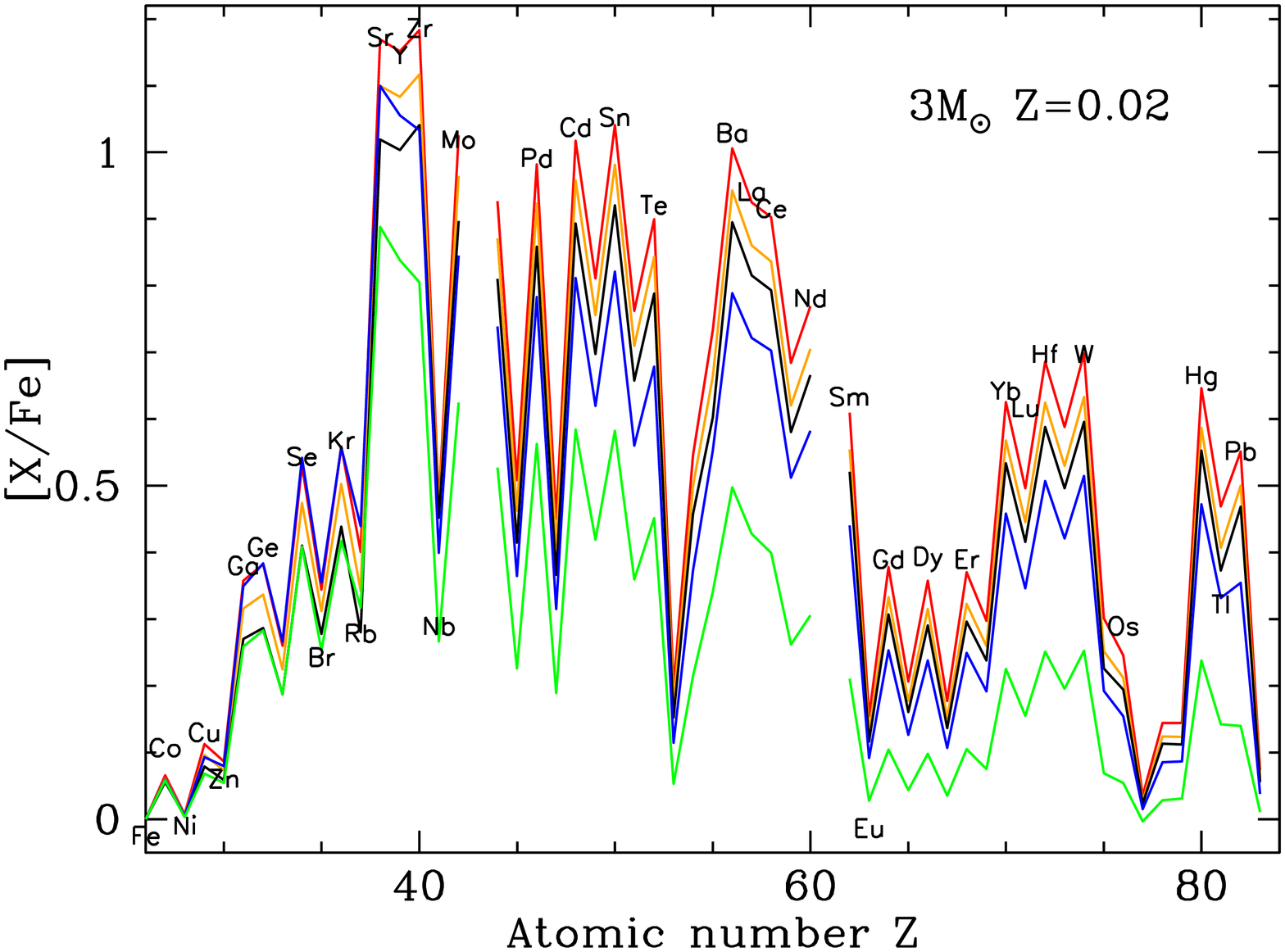} \\
\includegraphics[width=6.5cm,angle=-90]{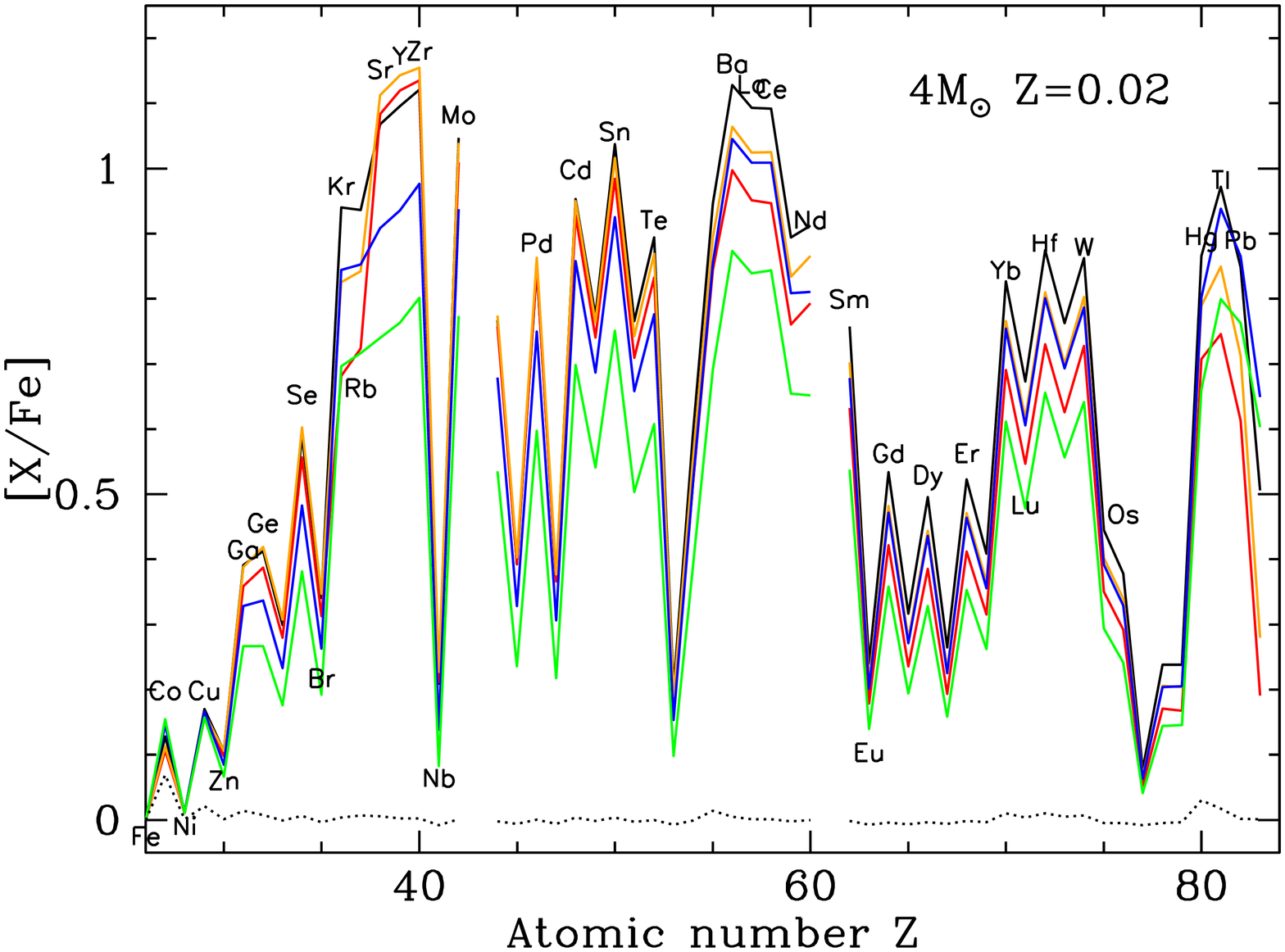} 
\includegraphics[width=6.5cm,angle=-90]{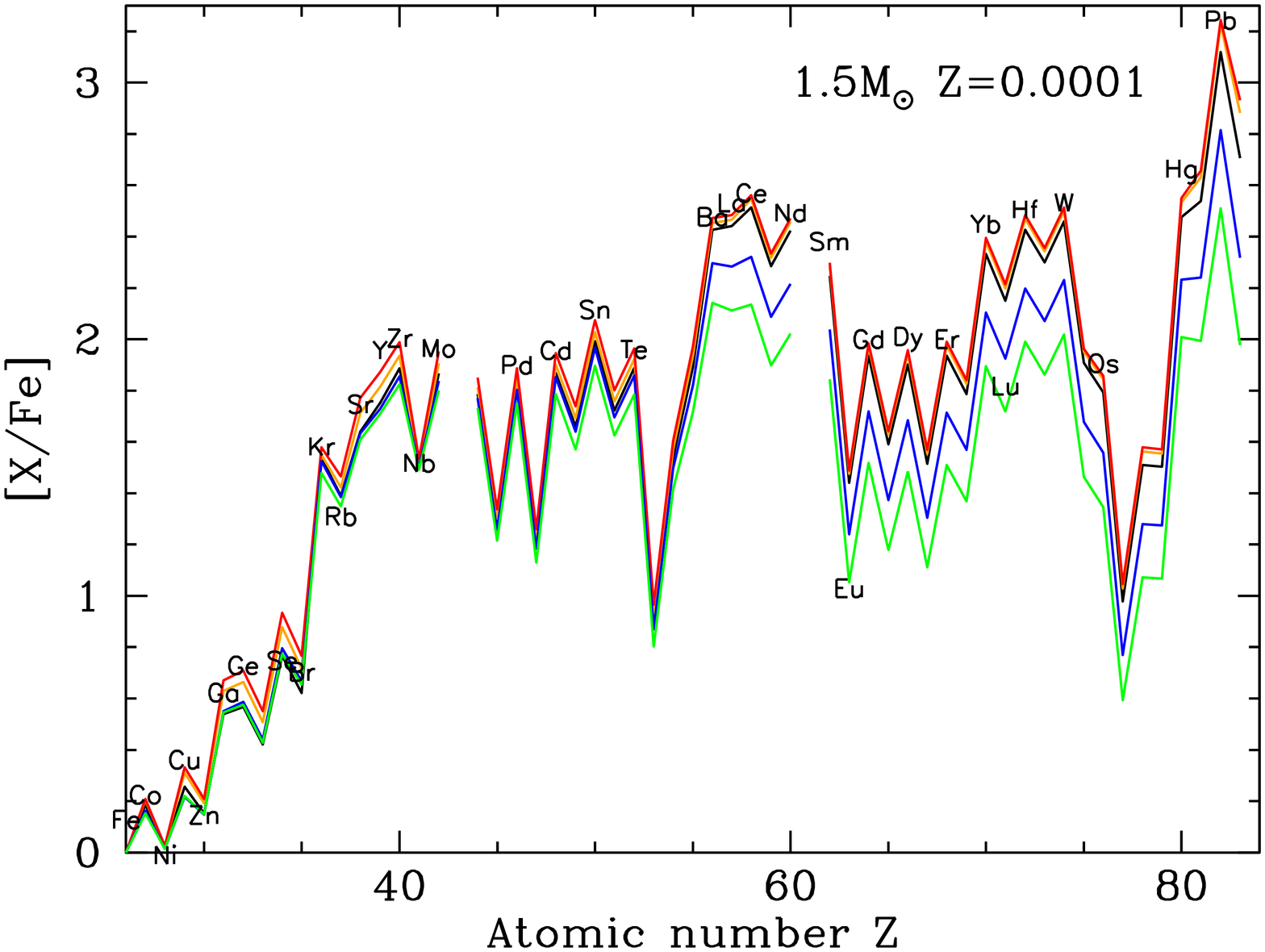} \\
\caption{[X/Fe] abundance ratios for the elements from Fe to Bi for the models of Set 1. 
The gaps in the distribution corresponds to the elements Tc and Pr, which have no stable 
isotopes. The black dotted line in the 4\,M$_\odot$ \emph{Z} = 0.02 panel represents the case where 
no PMZ was included. Note the change of scale for the y-axis between the panels.}
\label{fig:results1}
\end{figure*}

\begin{table}
\begin{center}
\centering
\caption{Selected final surface abundance ratios 
produced by all the stellar models of Set 1 and Set 2.
\label{table:abund}}
\setlength{\tabcolsep}{2.5pt}
\begin{tabular}{| l | *{5}{c} | }
\hline
{Profile} & {[Sr/Fe]} & {[Ba/Fe]} & {[Pb/Fe]} & {[Ba/Sr]} & {[Pb/Ba]}\\
\hline
\multicolumn{6}{c}{{\bf 1.25\,M$_\odot$ \emph{Z}=0.01}}\\
\hline 
$m^{3}$   & 0.31 & 0.02 & 0.01 & $-$0.29 & $-$0.01 \\
$m^{2}$   & 0.39 & 0.05 & 0.01 & $-$0.34 & $-$0.04 \\
$m$     & 0.57 & 0.25 & 0.01 & $-$0.32 & $-$0.24 \\
$m^{1/2}$ & 0.48 & 0.36 & 0.02 & $-$0.12 & $-$0.34 \\
$m^{1/3}$ & 0.40 & 0.40 & 0.03 & 0.00 & $-$0.37 \\ 
$-m$    & 0.39 & 0.05 & 0.01 & $-$0.34 & $-$0.04 \\
$-2m$    & 0.62 & 0.05 & 0.01 & $-$0.57 & $-$0.04  \\ 
$-3m$    & 0.03 & 0.01 & 0.01 & $-$0.02 & 0.00 \\
\hline
\multicolumn{6}{c}{{\bf 1.8\,M$_\odot$ \emph{Z}=0.01}}\\
\hline
$m^{3}$   & 1.09 & 1.38 & 0.83 & 0.29 & $-$0.55 \\
$m^{2}$   & 1.19 & 1.49 & 0.99 & 0.30 & $-$0.50 \\
$m$      & 1.24 & 1.62 & 1.15 & 0.38 & $-$0.47 \\
$m^{1/2}$ & 1.12 & 1.51 & 0.96 & 0.39 & $-$0.55 \\
$m^{1/3}$ & 0.95 & 1.28 & 0.66 & 0.33 & $-$0.62 \\
$-m$     & 1.37 & 1.76 & 1.39 & 0.39 & $-$0.37 \\
$-2m$    & 1.38 & 1.80 & 1.44 & 0.42 & $-$0.36 \\ 
$-3m$    & 1.07 & 0.50 & 0.07 & $-$0.57 & $-$0.43 \\
Test$^a$     & 1.08 & 0.67 & 0.06 & $-$0.41 & $-$0.61 \\
\hline
\multicolumn{6}{c}{{\bf 3.0\,M$_\odot$ \emph{Z}=0.01}}\\
\hline
$m^{3}$   & 1.35 & 1.67 & 1.31 & 0.32 & $-$0.36 \\
$m^{2}$   & 1.42 & 1.73 & 1.40 & 0.31 & $-$0.33 \\
$m$        & 1.46 & 1.68 & 1.38 & 0.22 & $-$0.30 \\
$m^{1/2}$ & 1.32 & 1.44 & 1.03 & 0.12 & $-$0.41 \\
$m^{1/3}$ & 1.10 & 1.17 & 0.64 & 0.07 & $-$0.53 \\
$-m$    & 1.57 & 1.91 & 1.72 & 0.34 & $-$0.19 \\
$-2m$    & 1.62 & 2.09 & 1.99 & 0.47 & $-$0.10 \\ 
$-3m$    & 1.42 & 1.17 & 0.62 & $-$0.25 & $-$0.55 \\
\hline
\multicolumn{6}{c}{{\bf 3.0\,M$_\odot$ \emph{Z}=0.02}}\\
\hline
$m^{3}$   & 1.17 & 1.01 & 0.55 & $-$0.16 & $-$0.46 \\
$m^{2}$   & 1.09 & 0.94 & 0.50 & $-$0.15 & $-$0.44 \\
$m$        & 1.02 & 0.89 & 0.47 & $-$0.13 & $-$0.42 \\
$m^{1/2}$ & 1.10 & 0.79 & 0.35 & $-$0.31 & $-$0.44 \\
$m^{1/3}$ & 0.89 & 0.50 & 0.14 & $-$0.39 & $-$0.36 \\
$-m$  & 1.44 & 1.46 & 1.20 & 0.02 & $-$0.26 \\
$-2m$    & 1.49 & 1.62 & 1.44 & 0.13 & $-$0.18 \\ 
$-3m$    & 0.89 & 0.39 & 0.05 & $-$0.40 & $-$0.34 \\
\hline 
\multicolumn{6}{c}{{\bf 4.0\,M$_\odot$ \emph{Z}=0.02}}\\
\hline
$m^{3}$   & 1.08 & 0.99 & 0.61 & $-$0.09 & $-$0.38 \\
$m^{2}$   & 1.11 & 1.06 & 0.71 & $-$0.05 & $-$0.35 \\
$m$       & 1.07 & 1.13 & 0.85 & 0.06 & $-$0.28 \\
$m^{1/2}$ & 0.91 & 1.04 & 0.86 & 0.13 & $-$0.18 \\
$m^{1/3}$ & 0.74 & 0.87 & 0.76 & 0.13 & $-$0.11 \\
$-m$     & 1.24 & 1.34 & 1.12 & 0.10 & $-$0.22 \\
$-2m$    & 1.38 & 1.55 & 1.46 & 0.17 & $-$0.09 \\ 
$-3m$    & 0.90 & 0.60 & 0.16 & $-$0.30 & $-$0.44 \\
\hline 
\multicolumn{6}{c}{{\bf 1.5\,M$_\odot$ \emph{Z}=0.0001}}\\
\hline
$m^{3}$   & 1.77 & 2.47 & 3.24 & 0.70 & 0.77 \\
$m^{2}$   & 1.71 & 2.48 & 3.23 & 0.77 & 0.75 \\
$m$  & 1.64 & 2.43 & 3.12 & 0.79 & 0.69 \\
$m^{1/2}$ & 1.63 & 2.29 & 2.81 & 0.66 & 0.52 \\
$m^{1/3}$ & 1.61 & 2.14 & 2.51 & 0.53 & 0.37 \\ 
$-m$     & 1.65 & 2.41 & 3.34 & 0.76 & 0.93 \\
$-2m$    & 1.61 & 2.37 & 3.53 & 0.76 & 1.16 \\ 
$-3m$    & 1.67 & 2.45 & 3.45 & 0.78 & 1.00 \\
\hline
\end{tabular}
\end{center}
$^a$Model run using the same proton profile as from Fig.~2 of \citet{trippella16}. 
\end{table}

The 1.8\,M$_\odot$ \emph{Z} = 0.01 model (top 
right panel of Fig.~\ref{fig:results1})
is the most representative of the case when the final abundance distribution is predominantly determined by 
radiative $^{13}$C burning. The neutron exposure in the \iso{13}C pocket in this model does not change significantly
(Table~\ref{table:tau}) and so do not the relative ratios [Ba/Sr] and [Pb/Ba], which are controlled by the neutron 
exposure, and are constant within $\pm$0.07 dex.
On the other hand, the absolute abundances vary by roughly a factor of two. They reach the 
maximum for the 
standard $m$ case, following the maxima of the $M_{\rm pocket}$
and the $M_{\rm tot}{\rm (^{\rm 13}C_{\rm eff})}$ 
(Table~\ref{table:c13pocketprop}). These results are qualitatively similar to those that would be obtained by 
varying the $M_{\rm PMZ}$ instead of the mixing profile: 
the neutron exposure, which controls the relative ratios, does not change; while the absolute amount of matter exposed
to the neutron flux, which controls the absolute ratios, changes.

The trend is similar in the 3\,M$_\odot$ \emph{Z} = 0.01 and \emph{Z} = 0.02 models and the 4\,M$_\odot$ 
\emph{Z} = 0.02 models. In these models -- even in the 4 \msun\ model, where the neutron density 
presents a complex
dependency on the mixing function (Fig.~\ref{fig:neutrondensity}) -- variations in the relative ratios 
[Ba/Sr] and [Pb/Ba] are typically lower than $\pm$0.2 dex. 
The most significant difference compared to the 1.8\,M$_\odot$ \emph{Z} = 0.01 model is that in the models 
of higher mass the \iso{22}Ne neutron source is also activated in the late TPs, which generally shift the 
abundance distribution towards the first $s$-process peak. An indirect effect from the PMZ on the activation
of the \iso{22}Ne results when a large \iso{14}N 
pocket ($m^{1/2}$ and $m^{1/3}$) is present, increasing the production of \iso{22}Ne. Ultimately this leads to a 
stronger neutron flux in the TP. 
Depending on the temperature in the TP it is possible to boost 
the production of Sr with respect to Ba, for example, in the 3\,M$_\odot$ \emph{Z} = 0.01 model; or 
the production of Ba and Pb with respect to Sr, for example, in the 4\,M$_\odot$ \emph{Z} = 0.02 model.
The final $s$-process distribution ultimately depends also on the number of TPs experienced by the model, 
which is higher in the higher mass models relative to the 1.8 \msun\ case. A higher number of TPs results
in a higher enrichment of the intershell, particularly of \iso{22}Ne 
in the cases where the \iso{14}N pocket is larger.  
More neutrons are captured by \iso{22}Ne inside the \iso{13}C pocket and 
the neutron exposure decreases.

In the case of the 1.5 \msun\ \emph{Z} = 0.0001 model, the decrease in the neutron exposure 
moving from the more shallow cases ($m^2$ and $m^3$) to the more steep ($m^{1/2}$ and $m^{1/3}$, 
Table~\ref{table:tau}) 
results in a decreased production of the second and third peak, with respect to the first peak. 

Finally, the 1.25 \msun\ model is very different from all the others because the neutron flux is dominated
by the \iso{13}C ingestions. Furthermore, 
the TDU is not efficient enough to produce a strong $s$-process signature at the 
stellar surface. Only the first peak elements are observed to be present 
in the $m^2$ and $m^3$ cases. In the other cases 
some production of Ba is achieved, while the third peak at Pb is never reached. This trend follows that
of the neutron exposure in the interpulse (Table~\ref{table:tau}). 
The enrichments achieved are not enough to explain the typical observation for AGB stars 
of roughly solar metallcity that [Ba/Fe]$\sim$1, 
hence, this model may be relevant only for specific applications, e.g., the 
production of \iso{86}Kr \citep{raut13} or, potentially, the 
composition of Sakurai's Object, as an alternative to the proton-ingestion episode \citep{herwig11}.

\subsubsection{Comparison of the results from Set 2}

\begin{figure*}
\centering
\includegraphics[width=6.5cm,angle=-90]{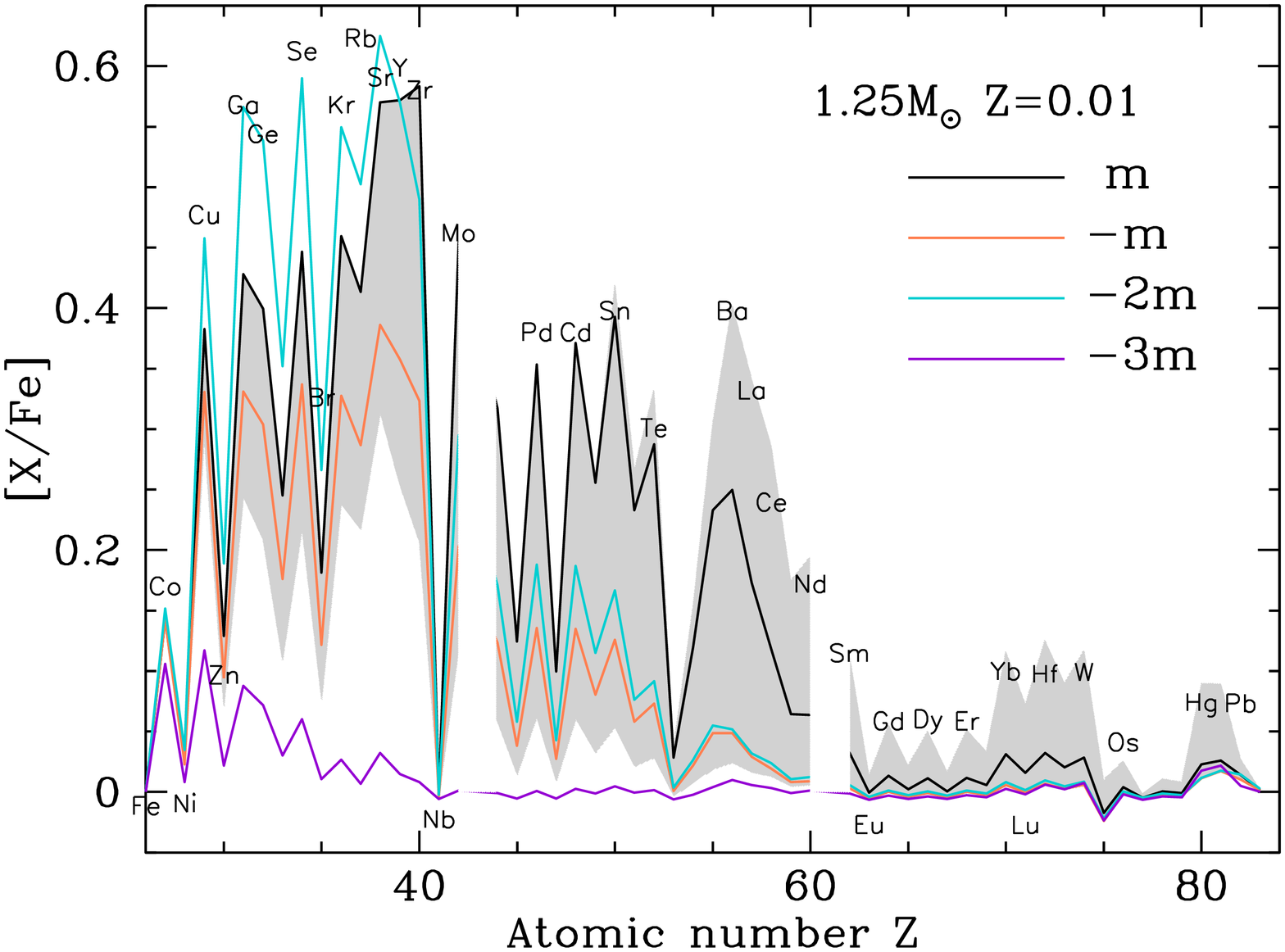} 
\includegraphics[width=6.5cm,angle=-90]{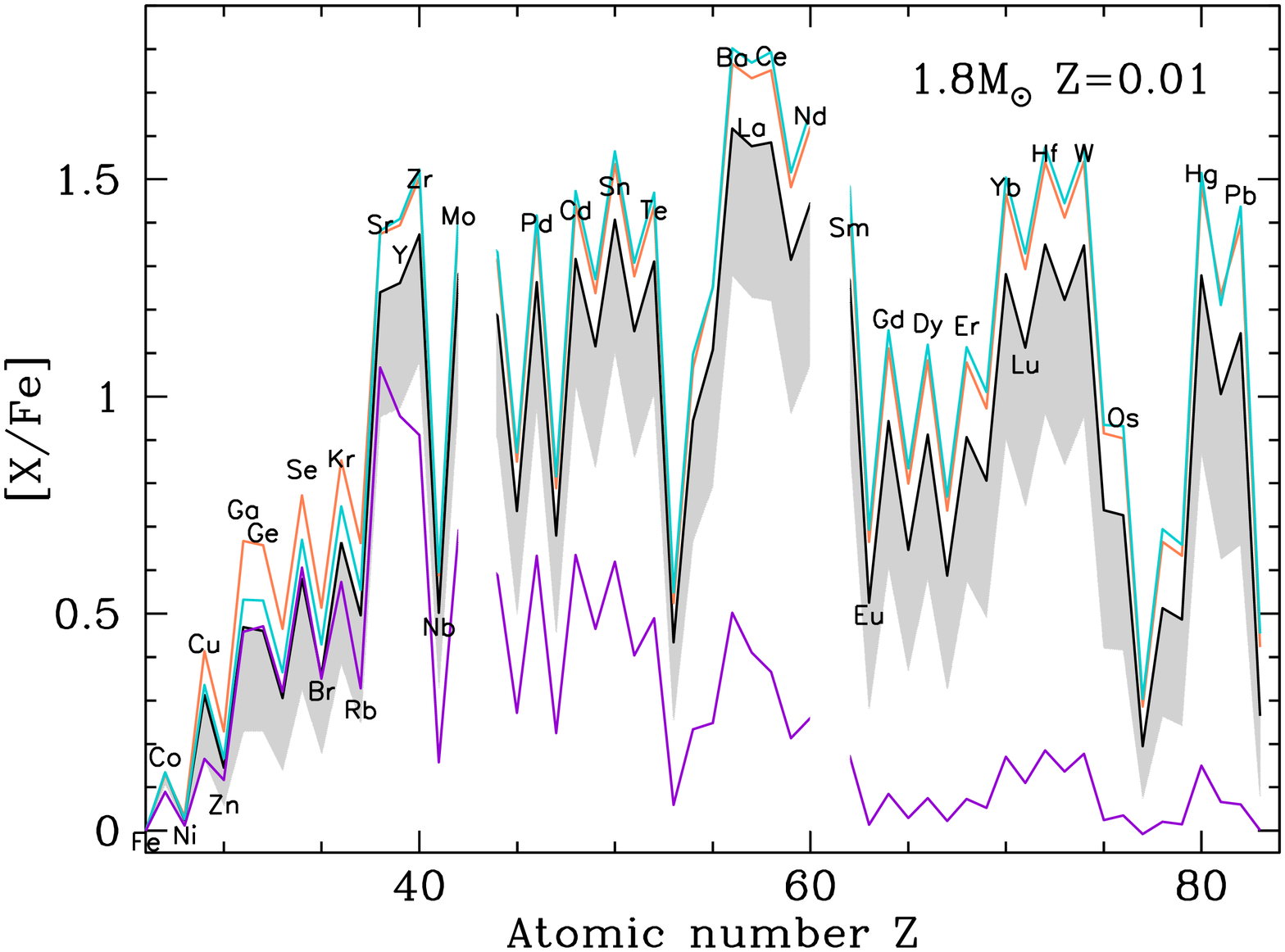} \\
\includegraphics[width=6.5cm,angle=-90]{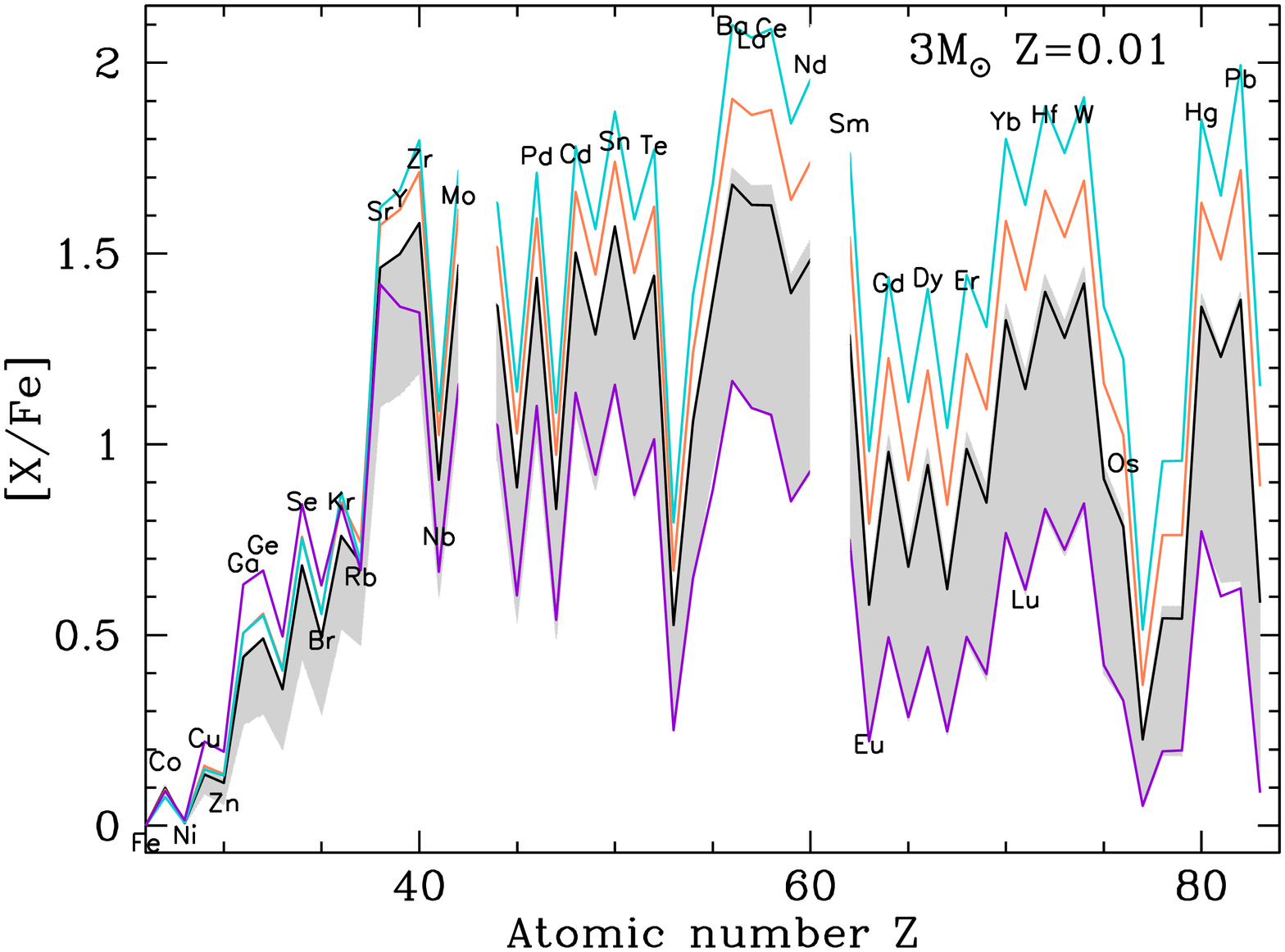} 
\includegraphics[width=6.5cm,angle=-90]{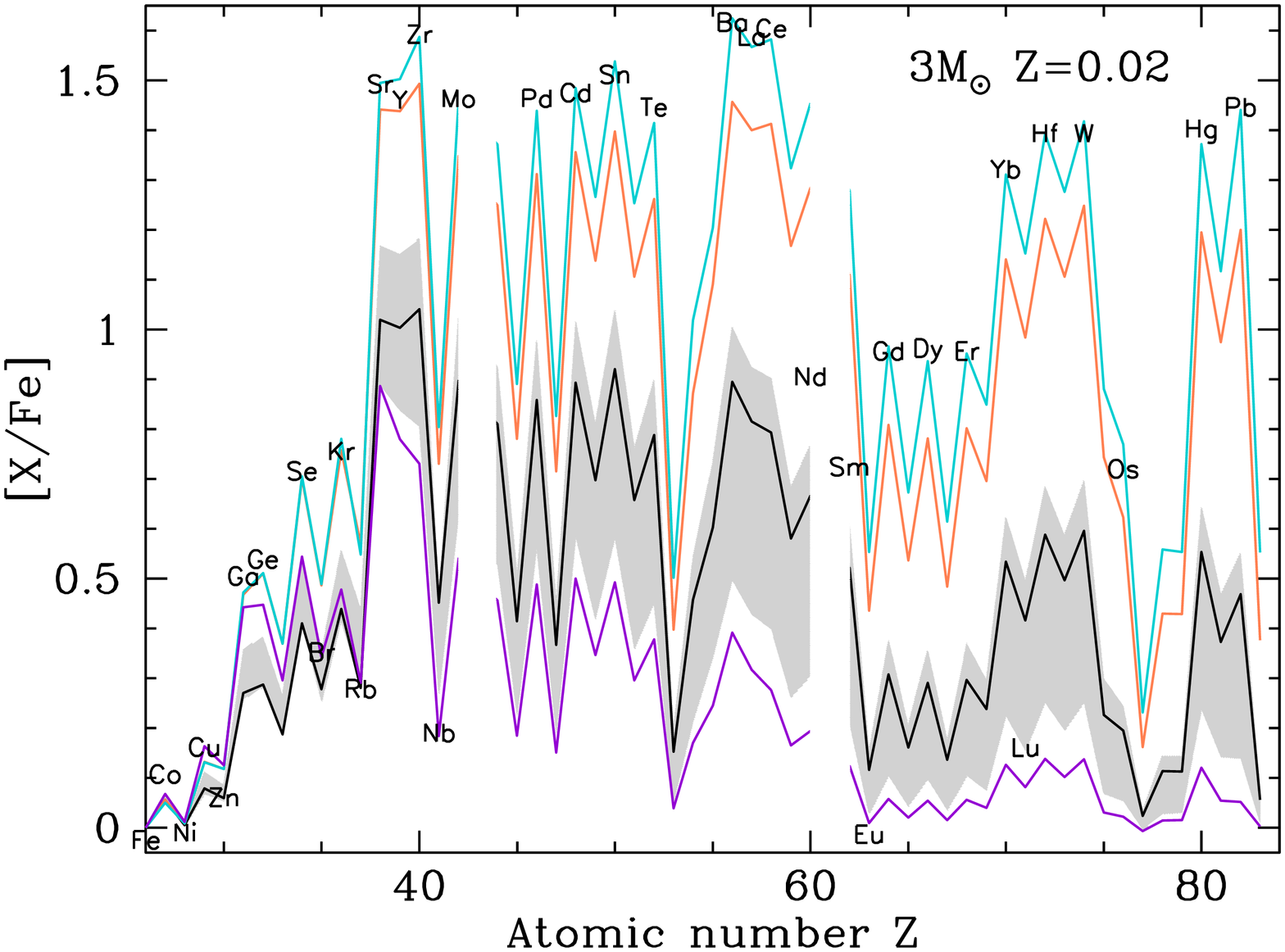} \\
\includegraphics[width=6.5cm,angle=-90]{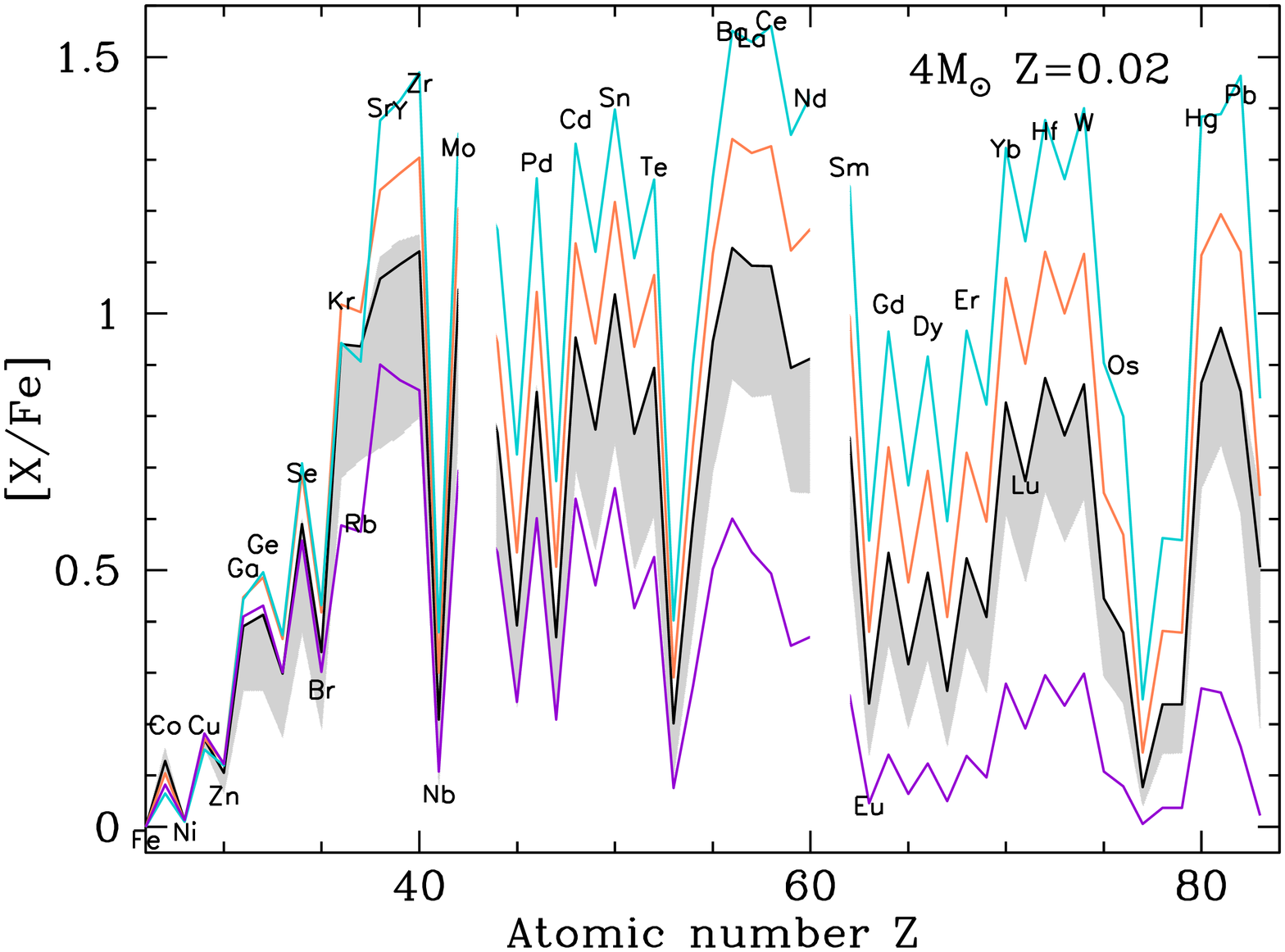}  
\includegraphics[width=6.5cm,angle=-90]{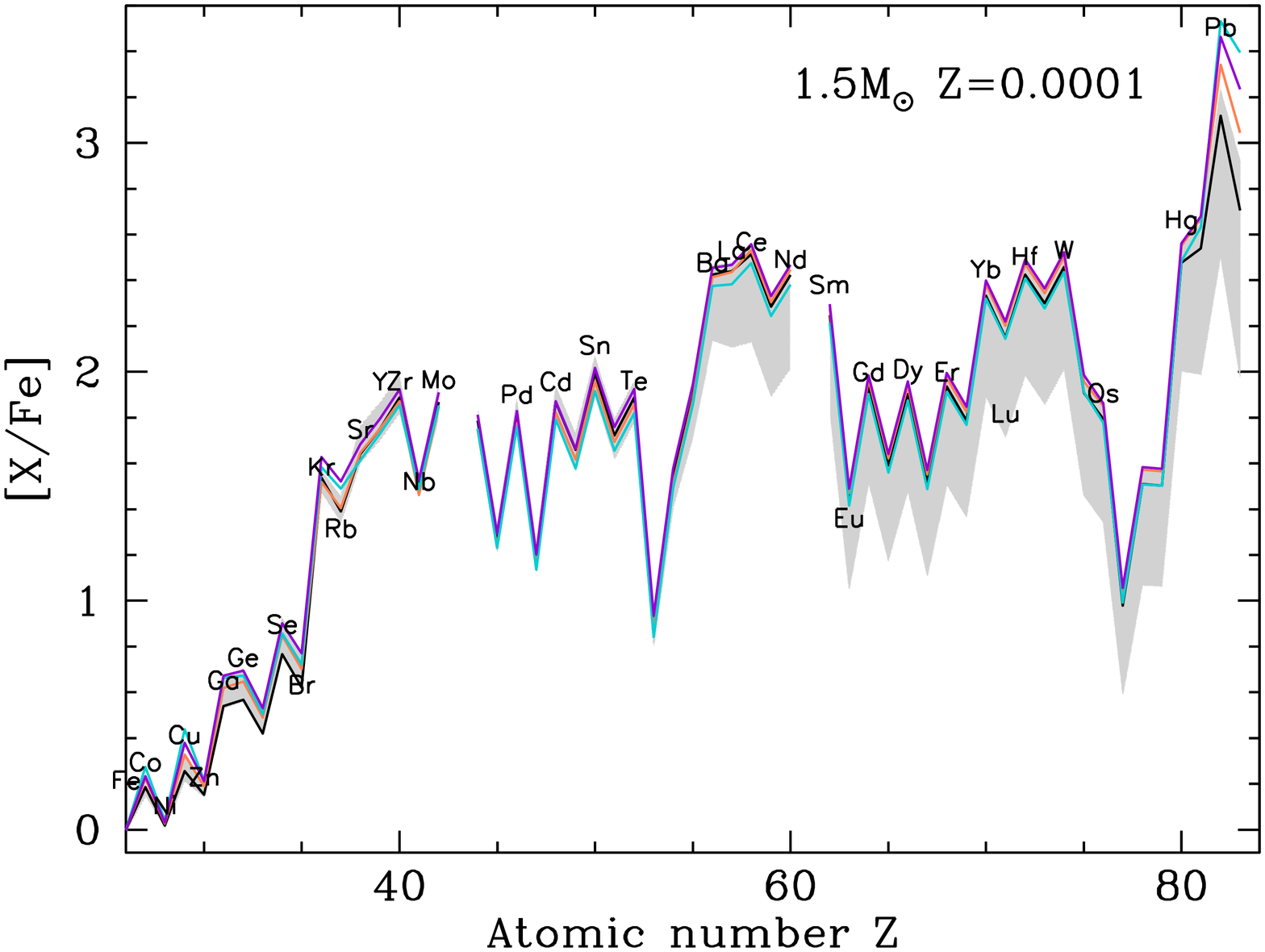} \\
\caption{Same as Fig.~\ref{fig:results1} but for the models from Set 2. For comparison, 
the grey shaded region represents the spread in abundance predictions from the 
Set 1 results.}
\label{fig:results2}
\end{figure*}

The cases of Set 2 generally present larger variations in the resulting abundances than the cases of 
Set 1. In the models of metallicity close to solar where the \iso{13}C pocket burns radiatively 
(1.8, 3, and 4 \msun) the main change in the results of the 
$-m$ and $-2m$ cases relative to the result of the 
standard $m$ case, is the increase of the absolute abundance production by up to $\sim$ 1 dex. 
On the other hand, the relative ratios only change  
within $\pm$0.1 dex. The reason is that, similarly to Set 1, the main effect of changing the 
profile in the way described by $-m$ and $-2m$ is 
to increase $M_{\rm pocket}$. Again, this is equivalent to increasing the total 
extent of the PMZ. The maximum of the absolute abundance follows the maximum $M_{\rm pocket}$ 
and $M_{\rm tot}{\rm (^{\rm 13}C_{\rm eff})}$, corresponding to case $-2m$.
The effect of the \iso{14}N pocket to increase the \iso{22}Ne abundance in the intershell 
is much milder 
in Set 2 than in Set 1 because the \iso{14}N pocket is a much smaller fraction of the PMZ 
(Table~\ref{table:c13pocketprop}). 

As in Set 1, the 1.25 \msun\ case characterised by the \iso{13}C ingestions behaves differently. 
The results for the $-m$ case are very similar to those from 
the $m^2$ case. 
Instead, in the $-2m$ case the peak of 
\iso{13}C is located at higher mass coordinate within the \iso{13}C pocket, i.e., at lower temperatures than 
those found deeper in the pocket. This means that effectively all the 
\iso{13}C in this case is ingested in the TPs (Fig~\ref{fig:profiles2}), 
which results in a lower neutron exposure and the production 
of the elements between Fe and Sr.  

With the $-3m$ case we enter a different regime. In this case the abundance of protons in the 
PMZ is below $10^{-3}$ all throughout the PMZ, hence the mass fraction of \iso{13}C is below 0.01 everywhere 
in the pocket (Fig.~\ref{fig:profiles2}).
The comparison of this case with the $m^{1/3}$ case from Set 1 illustrates very clearly 
why $M_{\rm tot}{\rm (^{\rm 13}C_{\rm eff})}$ cannot 
be used to uniquely constraint the final $s$-process distribution. 
In fact, $M_{\rm tot}{\rm (^{\rm 13}C_{\rm eff})}$ is very similar in the 
$-3m$ and $m^{1/3}$ cases, however, the final abundances and their relative patterns are 
completely different. This is due to way the \iso{13}C is distributed throughout the pocket 
(Fig.~\ref{fig:profiles2}). In the $m^{1/3}$ case, there is a peak in the abundance of \iso{13}C of
$\simeq$ 0.25, in mass fraction. In the $-3m$ case, instead, the maximum abundance of \iso{13}C 
is $\simeq$ 0.008, in mass fraction, i.e., roughly 3 times lower. In all the models where the \iso{13}C radiative 
burning is predominant, it is this {\em local} abundance of \iso{13}C that determines the neutron exposure 
and the final 
distribution. The abundances resulting from the 1.8, 3, and 4 \msun\ models, 
are generally the lowest in the $-3m$ case, 
and the relative distribution is shifted towards the first $s$-process 
peak. 

In the case of the 1.25 \msun\ model, the variations between the $-3m$ and the $m$ cases are even 
stronger than for the other models, probably due to the 
interplay between the rate of the \iso{13}C($\alpha$,n)\iso{16}O and of the 
\iso{14}N($\alpha$,$\gamma$)\iso{18}F reactions, the latter removing the neutron poison \iso{14}N, and the 
ingestion timescale, which is longer for the $-3m$ case, given that the \iso{13}C pocket 
is spread over a mass four times larger than in the $m^{1/3}$ case.

In the case of the 1.5 \msun\ model of low metallicity the effect is different. Because of the 
much lower number of Fe seeds, even the lowest concentration of \iso{13}C of case $-3m$ produces a 
significant neutron exposure, able to reach the third $s$-process peak. 

Due to the similarity of the $-3m$ proton profile to that produced by mixing driven by magnetic fields, 
to test the model uncertanties we further computed a 1.8 \msun\ $Z=0.01$ model (case Test in 
Table~\ref{table:abund}) using a polynomial fit of the profile shown in Fig.~2 of \citet{trippella16}. 
We find 
that the resulting ratios are extremely similar to those of the $-3m$ case. In half of the PMZ 
of \citet{trippella16} the proton abundance is lower than 10$^{-4}$. This results in 
a mass fraction of \iso{13}C lower than 10$^{-3}$, and a local neutron exposure lower than $\sim$0.05 
mbarn$^{-1}$ \citep[see Fig.~4 of][]{lugaro03a}. 
Such neutron exposure does not produce any significant amount of $s$-process elements and 
confirms that our approximation of setting the proton value to zero below 
10$^{-4}\,X_{\rm H}(CE)$ is valid. From Fig.~3 of 
\citet{trippella16}, one can derive [Ba/Sr]$\sim$0 (i.e., the solar system abundance distribution) and 
[Pb/Ba]$\simeq -0.3$ in the intershell at the end of the evolution of a 1.5 \msun\ star of [Fe/H]=$-$0.15. 
These values are different from those that we obtain at the end of the evolution in the intershell: 
[Ba/Sr]=$-$0.38 and [Pb/Ba]=$-$0.97 (essentially no Pb is produced in our 
model\footnote{Note that at the stellar surface we obtain instead 
[Ba/Sr]=$-$0.41 and [Pb/Ba]=$-$0.61 (Table~\ref{table:abund}).
The difference between the intershell and the surface [Pb/Ba] values  
is due to the signature of the 
envelope matter of solar distribution in the surface value, which brings the [Pb/Ba] ratio closer to 
zero.}). 
The origin of these differences is unclear since on top of implementing exactly the same 
proton profile we have used the same metallicity. The initial stellar 
masses are very close to each other (1.8 versus 1.5 \msun) and low enough to ensure that the \iso{22}Ne
neutron source does not play any role. 
One possibility is the different number of PMZ inserted: 5 in our model and 
9 in the case of \citet{trippella16}. However, the intershell abundance distribution in our model has 
reached an asymptotic limit by the end of the computed evolution: at the previous interpulse period
the [Ba/Sr] and [Pb/Ba] ratios are within 0.15 dex of their final values 
and we do not expect any large variations if the evolution was extended. 

\section{Discussion}

\begin{figure}
\begin{center}
\includegraphics[width=8.5cm]{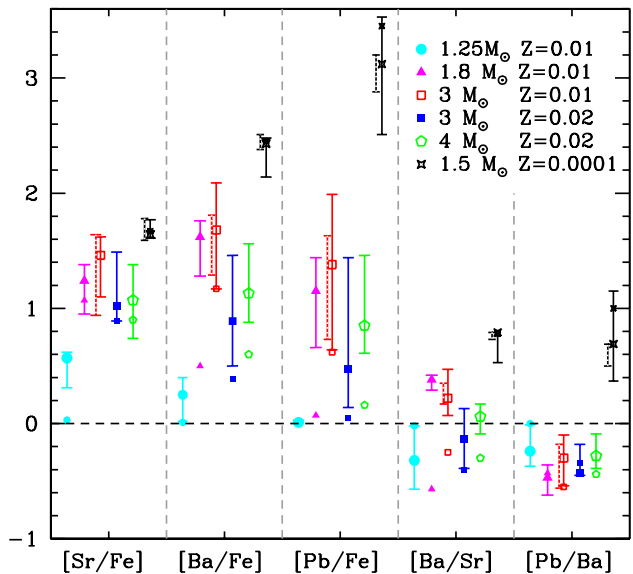}
\end{center}
\caption{Summary of the selected final abundance ratios for the different stellar models. 
The large symbols represent the standard case $m$, and the solid-line 
error bars represent the spread in values from all the models of Set 1 and Set 2, 
except for the $-3m$ model,  
plotted instead as a smaller symbol. Also shown are the ratios obtained 
with the standard $m$ mixing profile but different values of 
$M_{\rm PMZ}$ (dashed error bars) for the 3 \msun\ $Z=0.01$ model (with $M_{\rm PMZ}$ from $5 
\times 10^{-4}$ to $4 \times 10^{-3}$ \msun) and for the 1.5 \msun\ $Z=0.0001$ model 
\citep[with $M_{\rm PMZ}$ from $6 \times 10^{-4}$ to $4 \times 10^{-3}$ \msun, from][]{lugaro12}.}
\label{fig:resultsfinal}
\end{figure}

The $s$-process results from our stellar models calculated with a variety of mixing profiles 
leading to the formation of the \iso{13}C pocket are
summarised in Fig.~\ref{fig:resultsfinal}. A main finding is that for the models of metallicity around 
solar in which the 
predominant neutron flux is produced by the \iso{13}C pocket burning radiatively, the effect of 
changing the mixing profiles is in first approximation 
equivalent to changing the mass extent of the PMZ, $M_{\rm 
PMZ}$. The absolute abundances increase with the extext of the \iso{13}C pocket, $M_{\rm pocket}$, e.g., 
the $-2m$ case dominates the upper limits of the absolute abundance ratios, while the 
$m^{1/3}$ case dominate the lower limits. For the $-3m$ case the abundances are even lower
and are plotted 
as a separate symbol in Fig.~\ref{fig:resultsfinal}. Variations reach up to roughly $\pm$0.5 dex.
On the other hand, the 
relative ratios do not change within $\pm$0.2 dex. This result is summarised and illustrated by 
the 1.8, 3, and 4 \msun\ models shown in Fig.~\ref{fig:resultsfinal}, together with the 
results obtained for the 3 \msun\ $Z=0.01$ model calculated with the standard $m$ mixing profile but 
different values of $M_{\rm PMZ}$: the range of variation is the same when changing the 
mixing profile and when changing $M_{\rm PMZ}$. 

The 1.25 \msun\ $Z=0.01$ is different from the others in that the effect of the mixing profile is 
much more limited: both absolute and relative ratios vary at most by $\pm$0.2 dex. In this model
the prominent effect is the ingestion of the \iso{13}C pocket, which somewhat suppresses 
the neutron flux and produces a different abundance pattern and milder $s$-process signatures.
According to \citet{bazan93}, feedback effects from the ingestion onto the stellar structure 
would lead to higher neutron density and lower neutron exposures than what we have calculated here.
The abundances pattern would depart from that resulting from 
the \iso{13}C burning radiative cases in the same way as we have found here, but to a larger extent.

In the low-metallicity 1.5 \msun\ $Z=0.0001$ model instead variations due to changing the 
mixing profile are larger than those obtained by changing the value of $M_{\rm PMZ}$ 
(Fig.~\ref{fig:resultsfinal}). The most significant effects are seen on Pb because in 
this model the neutron exposure is high enough to produce fixed, 
equilibrium abundances of Sr and Ba. The [Pb/Ba] 
ratio varies by almost an order of magnitude, with the highest value (1.16) 
achieved for the $-2m$ profile
and the lowest (0.37) with the $m^{1/3}$ profile. This makes variations in the mixing profile 
at low metallicity one of the possible factors that could contribute to explaining 
the variations in [Pb/Ba] (within a similar 
range as found here) observed in carbon-enhanced metal-poor (CEMP) stars 
with enhancements in the $s$-process elements
\citep{vaneck03,bisterzo10,bisterzo11,bisterzo12,lugaro12}.

One problem related to CEMP stars is that the standard $m$ models 
that produce [Pb/Ba]$\sim$1 
typically result in [Na/Ba] ratios more than an order of magnitude higher than observed
\citep[see Fig.~9 of][]{lugaro12}.
Among the elements lighter than Fe, Na is the only one whose abundance 
changes significantly for the different profiles tested here (Table~\ref{table:Na}). 
This is because  
\iso{23}Na (the only stable isotope of Na) is produced in the H-burning shell and in the top layers of the 
PMZ by proton captures on 
\iso{22}Ne and in the presence of neutrons in the \iso{13}C pocket and in the TPs via neutron captures 
on \iso{22}Ne \citep{goriely00,cristallo09,bisterzo11,lugaro12}.
As a consequence both the mixing 
profile itself and its effect on the intershell abundance of \iso{22}Ne play a 
significant role. 
Interestingly, the $-2m$ and $-3m$ cases produce the highest [Pb/Ba] and the lowest [Na/Ba] of 
all models. However, even the lowest value we find ($-$0.90) is still higher than the observations 
(down to $-$2 dex). \citet{cavanna15} reported 
a revised value of the \iso{22}Ne(p,$\gamma$)\iso{23}Na rate measured underground, however, at the 
temperature of interest ($<$ 20 MK) the new rate is the same as the 
rate we have used here \citep{iliadis10}.
On the other hand, the rate of the destruction reaction 
\iso{23}Na(p,$\alpha$)\iso{20}Ne is poorly known \citep{iliadis10}. While the 
issue of the nuclear physics input is still open and needs further investigation, 
observation of an anticorrelation between the Pb and the Na abundance in CEMP stars
may strengthen the possibility that the Pb and Na variations are indeed due to variations in the mixing 
profile of the PMZ. 

\begin{table}
\centering
\caption{The [Na/Fe], [Na/Ba], and [Pb/Ba] ratios for the 1.5\,M$_\odot$ \emph{Z} = 
0.0001 model and the cases of Set 1 and Set 2.}
\setlength{\tabcolsep}{2.5pt}
\begin{tabular}{lccccccccc}
\hline
 & $m^3$ & $m^2$ & $m$ & $m^{1/2}$ & $m^{1/3}$ & $-m$ & $-2m$ & $-3m$ \\
\hline
{[Na/Fe]} & 1.86 & 2.01 & 2.28 & 2.54 & 2.66 & 1.91 & 1.61 & 1.55 \\
{[Na/Ba]} & $-$0.61 & $-$0.47 & $-$0.15 & 0.25 & 0.52 & $-$0.50 & $-$0.76 & $-$0.90 \\
{[Pb/Ba]} & 0.77 & 0.75 & 0.69 & 0.52 & 0.37 & 0.93 & 1.16 & 1.00 \\
\hline
\end{tabular}
\label{table:Na}
\end{table}

\subsection{Comparison with meteoritic stardust data}

Stardust grains have been recovered from meteories for the past three decades, and have been used to 
constrain models of nucleosynthesis processes in the stars from where they originated 
\citep{zinner14}. The vast majority of these grains formed in the expanding envelopes of AGB stars, in 
particular, silicon carbide (SiC) grains formed in C-rich AGB stars, since C$>$O is a necessary condition 
for the formation of SiC molecules. They carry a strong signature of $s$-process nucleosynthesis in the 
isotopic composition of elements heavier than Fe that are present in trace amounts. 
Because such compositions can be measured to 
very high precision, particularly when using the Resonant Ionisation Mass Spectromety (RIMS) technique 
\citep[e.g.][]{liu14a,liu14b,liu15}, stardust SiC grains 
provide excellent constraints for stellar models of 
the $s$ process. On the other hand, we do not know {\it a priori} the initial mass and metallicity of 
the parent star of each grain. The general constraints, also coming from the composition of the light 
elements C, N, Ne, and Si, are that the SiC parent stars should be of metallicity roughly around solar, 
and should have initial masses below roughly 5 \msun\ \citep{lugaro99,lugaro03b}. 

\begin{figure*}
\centering
\includegraphics[width=6.5cm,angle=-90]{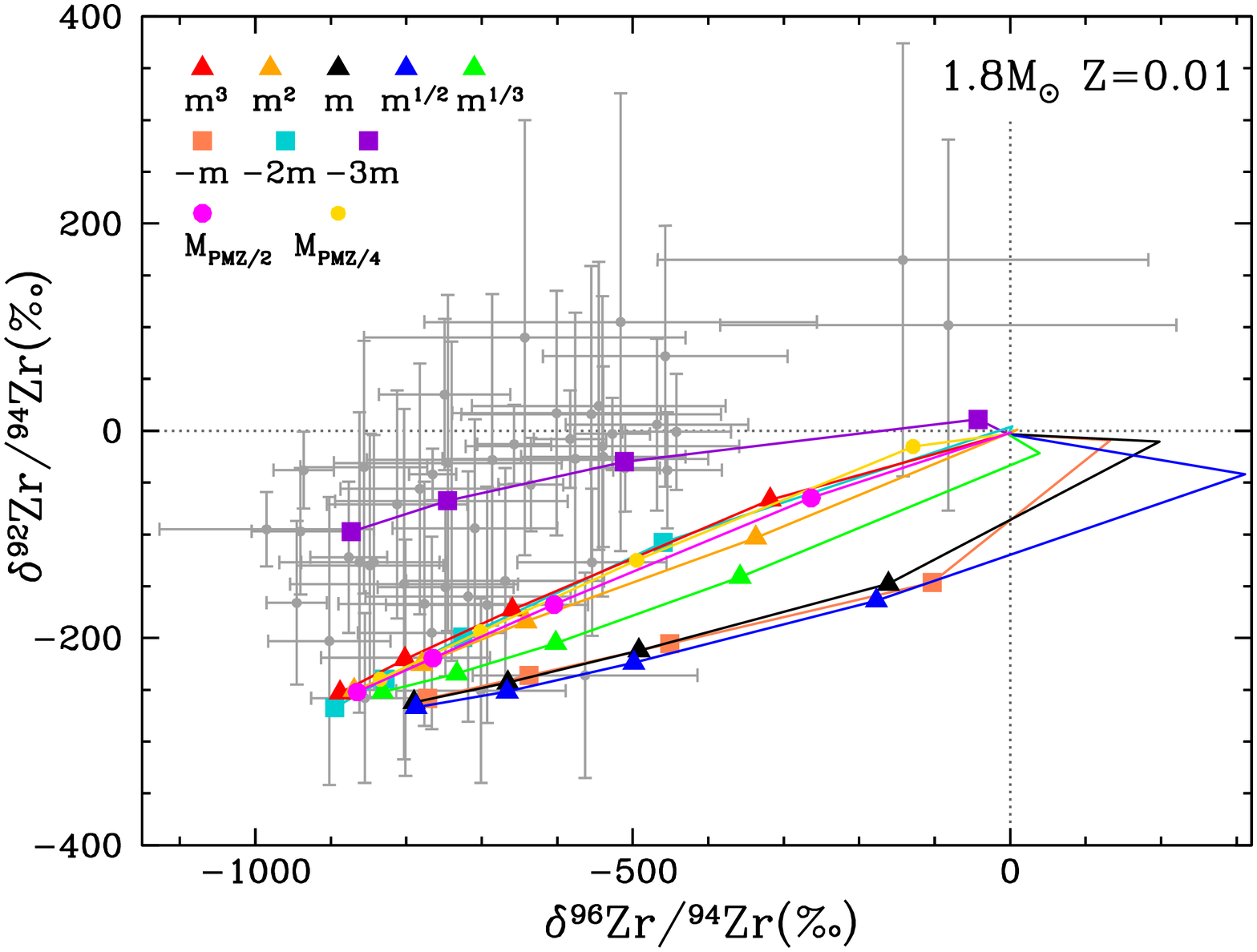} 
\includegraphics[width=6.5cm,angle=-90]{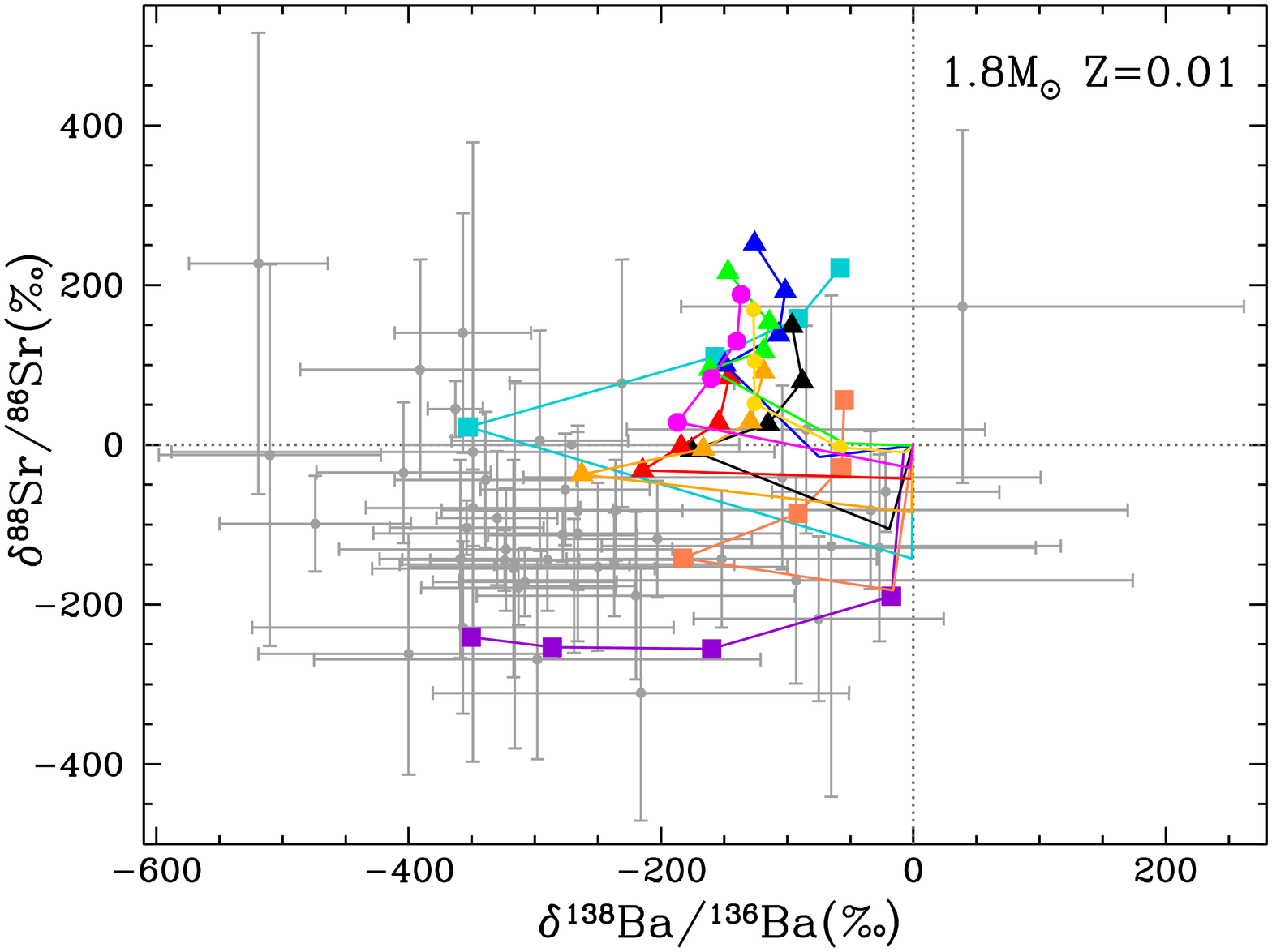} \\
\includegraphics[width=6.5cm,angle=-90]{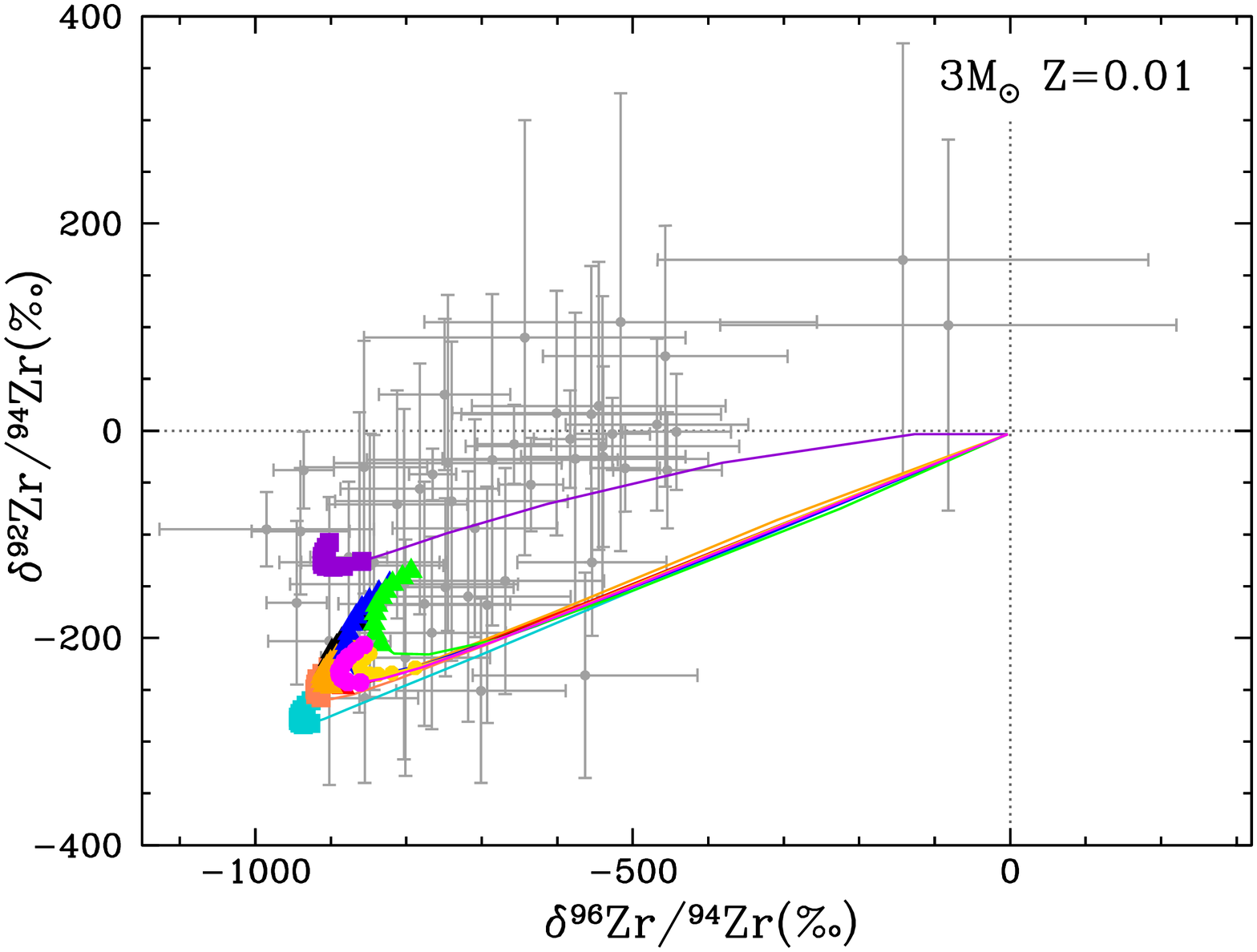} 
\includegraphics[width=6.5cm,angle=-90]{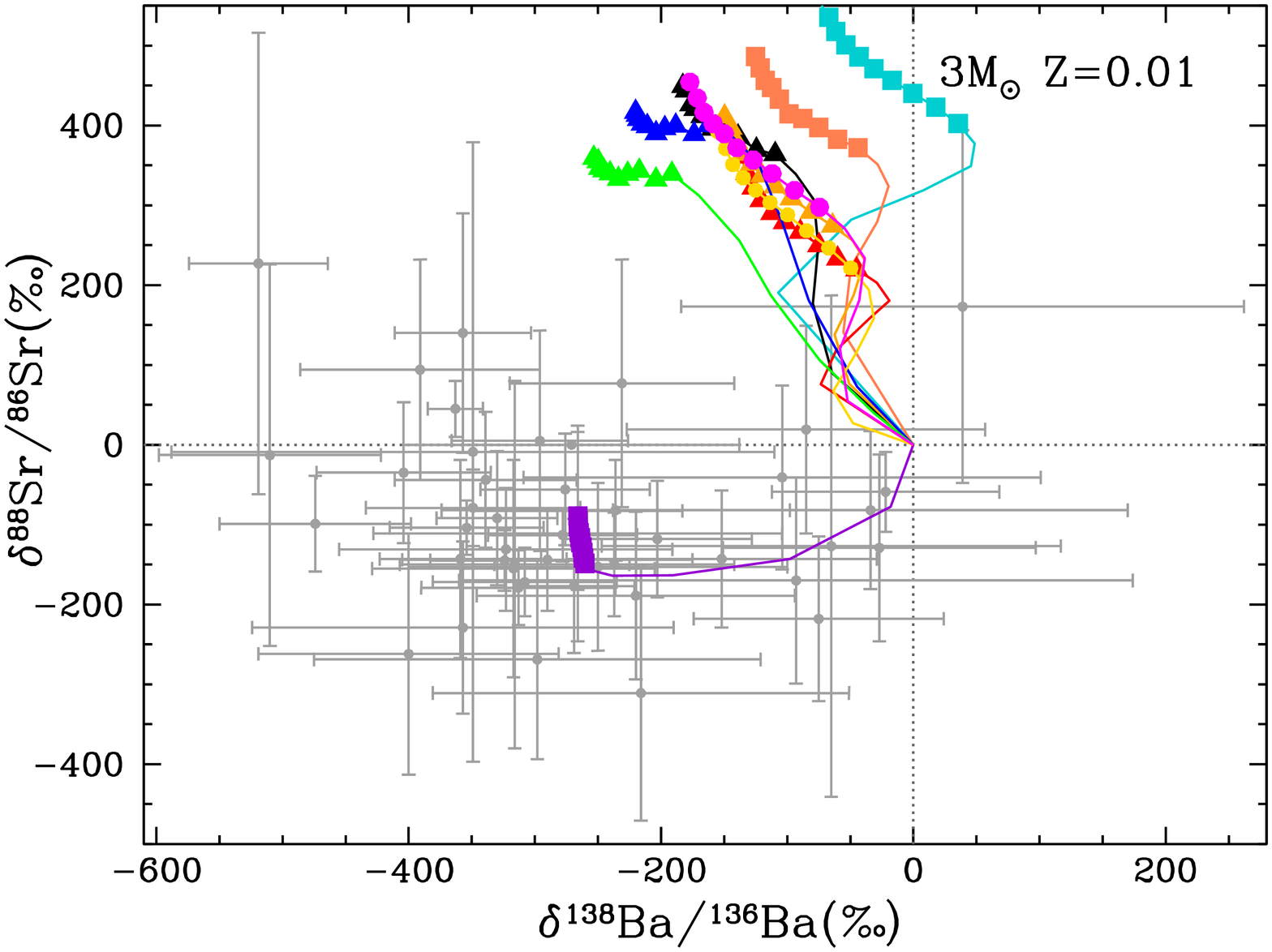} \\
\includegraphics[width=6.5cm,angle=-90]{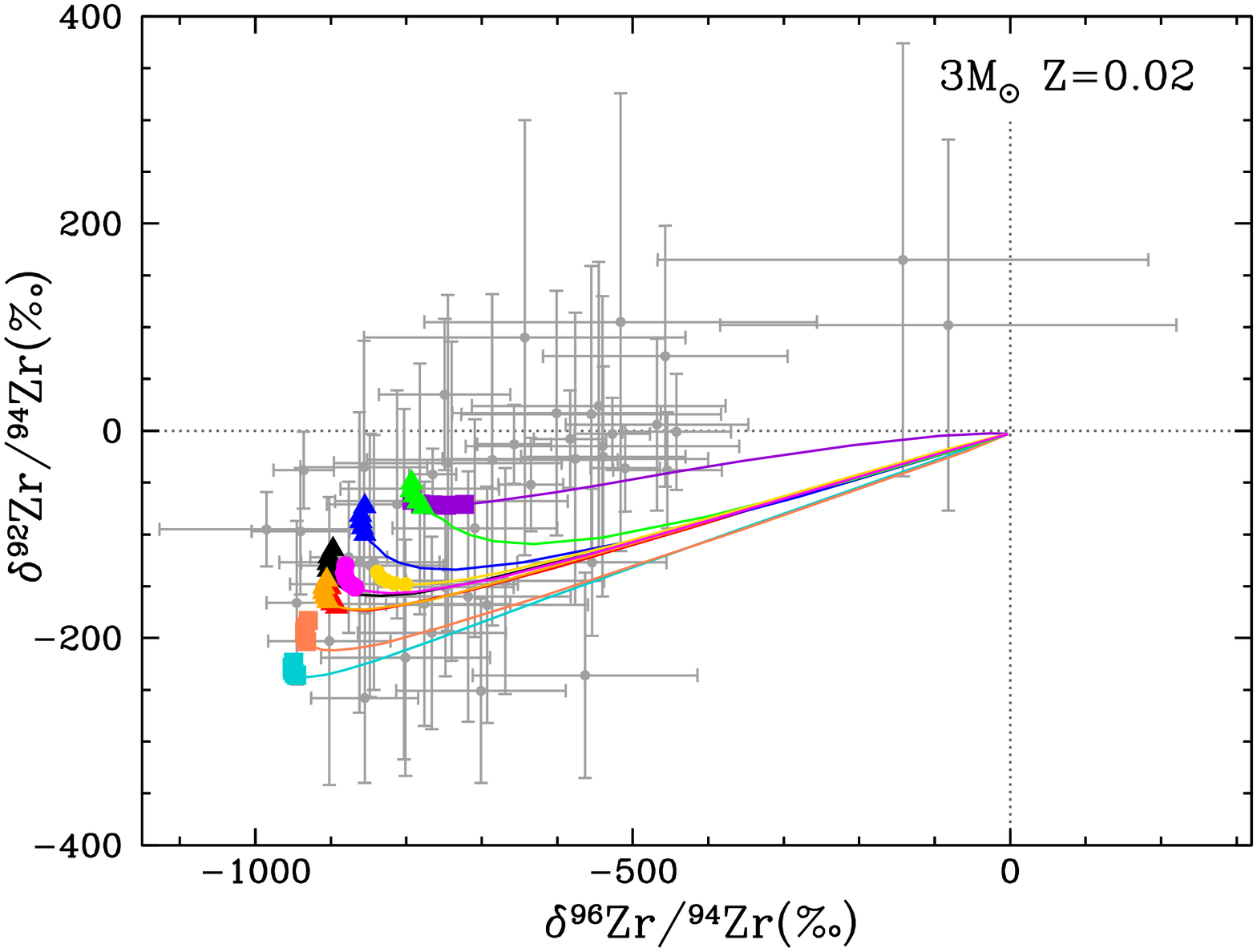}  
\includegraphics[width=6.5cm,angle=-90]{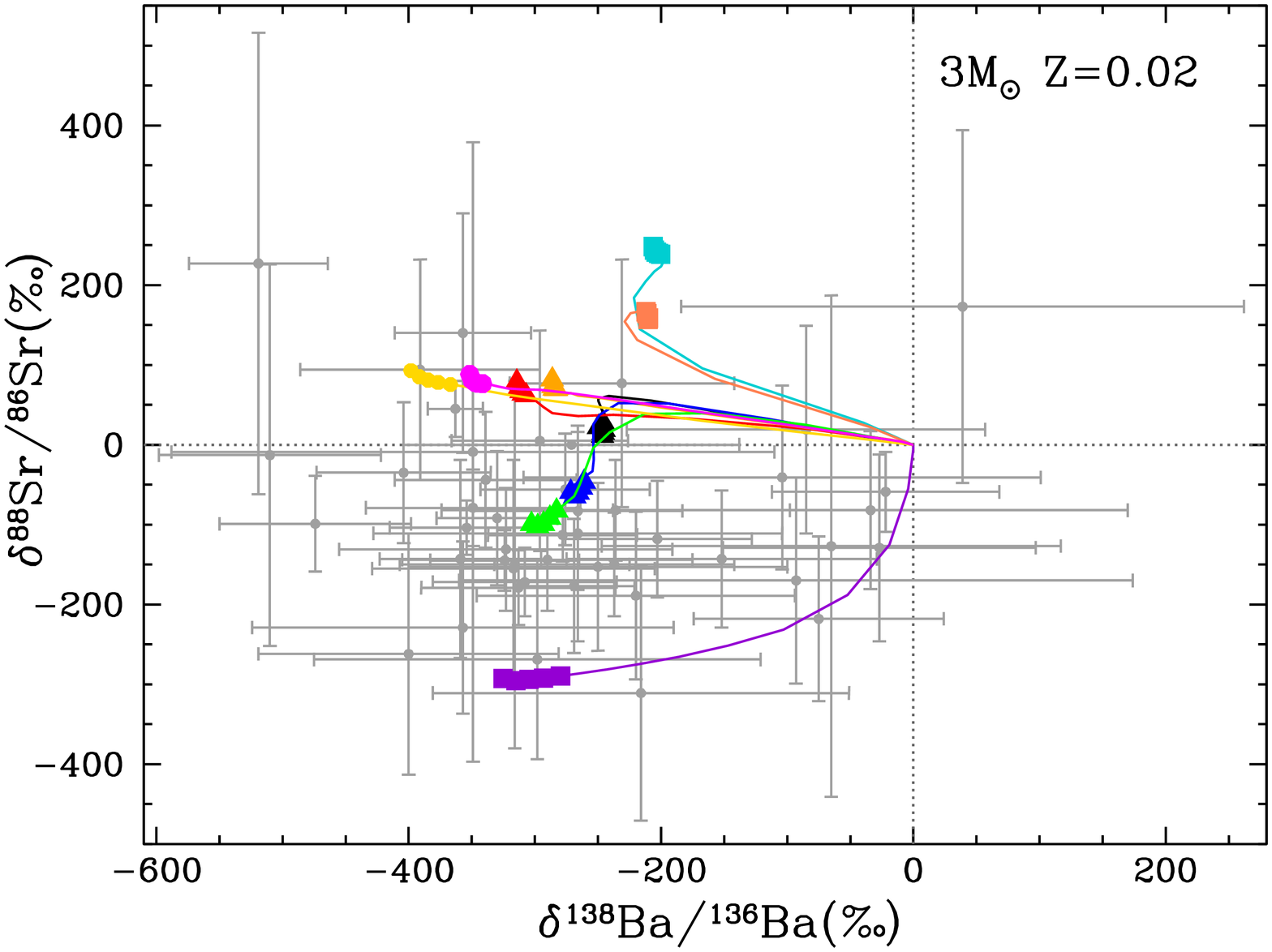} \\
\caption{A selection of our models 
of close to solar metallicity for all the different mixing profile cases is compared to the 
$s$-process composition observed in stardust SiC grains from AGB stars (grey symbols with 2$\sigma$ 
error bars). We use the standard $\delta$ notation, which represent permil variations of the indicated 
isotopic ratio with respect to the solar value (with $\delta=0$, by definition). The colored lines show 
the evolution at the stellar surface starting from the solar initial ratios. The overplotted
solid symbols correspond to TDU episodes with a C-rich envelope, the 
necessary condition for the formation of SiC. The triangles correspond to Set 1 and the squares to Set 
2, with the same colors as in all the previous plots. The circles represent the cases where $M_{\rm PMZ}$ 
was decreased as indicated.}
\label{fig:grains}
\end{figure*}

In Fig.~\ref{fig:grains} we compare a selection of our models with 
the composition of the grains. We choose isotopic ratios that are sensitive to the neutron 
exposure, hence to the features of the \iso{13}C pocket, as well as the mass and metallicity of the 
star. As noted above for the elemental ratios, also in terms of the isotopic ratios, typically 
varying the mixing 
profile produces very similar results as varying its extent in mass, except for the $-3m$ case. Only for 
the 3 \msun, $Z=0.02$ there are significant differences, where the variations in the mixing profile 
produces a larger variety of isotopic compositions than variations in $M_{\rm PMZ}$.

The \iso{88}Sr/\iso{86}Sr ratio is an 
excellent tracer of the neutron exposure. It is clear from Fig.~\ref{fig:grains} that the grains require 
a neutron exposure lower than that produced using the majority of the mixing profiles and stellar models 
tested here. There are several ways to 
achieve this: one is to use the $-3m$ mixing profile, which as discussed above stands out from the 
others for producing a markedly lower neutron exposure. Another possibility is to increase the stellar 
metallicity, i.e., when comparing the 3 \msun, $Z=0.01$ to the 3 \msun, $Z=0.02$ model, it is clear that 
the latter provides a better match to the grains, also for the $m^{1/2}$ and $m^{1/3}$ mixing profiles. 
In fact, \citet{lugaro17} demonstrated that AGB stars of
metallicity around $Z=0.03$, i.e., roughly twice higher than solar, produce most of 
the signatures observed in the grains for Sr, Zr, and Ba.

In terms of the Zr isotopic ratios, similar conclusions 
can be drawn: the majority of the models with metallicity lower than solar ($Z=0.01$) can only match the 
lowest \iso{92}Zr/\iso{94}Zr ratio observed in the grains. Lower neutron exposures are required to 
reach the observed average, i.e., again either using the $-3m$ case, or a higher stellar metallicity. 
However, only the 
specific case of the $-3m$ profile in the 1.8 \msun, $Z=0.01$ can reach  
\iso{92}Zr/\iso{94}Zr ratios around solar together with  
\iso{96}Zr/\iso{94}Zr ratios roughly 50\% lower than solar, as measured in many grains.

\section{Conclusions}

We have found that varying the mixing profile of the PMZ generally produces the same
results as varying the extent of the PMZ. This means that in the vast majority of 
$s$-process producing AGB stars, those where \iso{13}C burns radiatively: 
\begin{enumerate}
\item{the mixing profile and the extent of the PMZ can be 
considered as the same free parameter, unless the profile is changed to the extreme extent of case $-3m$;}
\item{when the stellar mass is above $\sim$2 \msun, the changes are 
dominated by the feedback of the mixing profile on the operation 
of the \iso{22}Ne neutron source, rather than of the \iso{13}C neutron source itself;}
\item{the overal effect on the relative $s$-process 
distribution is of the order of a factor of two, comparable to typical spectroscopic 
observational error bars;}
\item{the Sr, Zr, and Ba composition observed in stardust SiC grains requires lower neutron 
exposures than those experienced in the majority of our models, the $-3m$ profile is the most promising 
case, although it cannot match simultaneously 
all the different ratios.}
\end{enumerate}
The low-metallicity AGB model is an exception, where the [Pb/Ba] ratio is significantly 
affected, with the possible interesting consequences for the Pb abundance in CEMP stars discussed above.
For all the other cases, the abundance 
of \iso{12}C in the intershell, which is potentially enhanced by overshoot into the core, 
the presence of \iso{13}C ingestions, which are also affected by the uncertain 
rate of the \iso{13}C($\alpha$,n)\iso{16}O 
reaction \citep{guo12}, and the effect of 
rotational mixing appear to represent more significant uncertainties for the $s$ process than the details
of the mixing function leading to the formation of the PMZ. 
To match the composition of the stardust SiC grains further potential effects due to their 
parent stars being of 
different masses and metallicity than considered here need to be taken into account \citep{lugaro17}.

\section{Acknowledgments}
We thank the anonymous referee for important comments, particularly for pushing us to include Section 
5.1 on the comparison with stardust grains.
M.~L. is a Momentum (``Lend\"ulet-2014'' Programme) project leader of the Hungarian
Academy of Sciences. 
This research was supported under 
the Go8-DAAD Australia/Germany Joint research cooperation scheme. 
R.~J.~S. is the recipient of a Sofja Kovalevskaja Award from the Alexander von Humboldt Foundation.
\\

\bibliographystyle{mn2e} 
\bibliography{ref} 

\end{document}